\documentclass[]{spie}  %>>> use for US letter paper
\usepackage{amsmath,amsfonts,amssymb}
\usepackage{graphicx}
\usepackage[colorlinks=true, allcolors=blue]{hyperref}

\usepackage{booktabs} % Horizontal rules in tables
\usepackage{comment}

\title{LUVOIR-ECLIPS closed-loop adaptive optics performance and contrast predictions}

\author[a]{Axel Potier}
\author[a]{Garreth Ruane}
\author[a]{Pin Chen}
\author[b]{Ankur Chopra}
\author[b]{Larry Dewell}
\author[c]{Roser Juanola-Parramon}
\author[b]{Alison Nordt}
\author[d]{Laurent Pueyo}
\author[a]{David Redding}
\author[a]{A.J. Eldorado Riggs}
\author[e]{Dan Sirbu}

\affil[a]{Jet Propulsion Laboratory, California Institute of Technology, 4800 Oak Grove Drive, Pasadena, CA 91109}
\affil[b]{Lockheed Martin Space, Advanced Technology Center, 3251 Hanover Street, Palo Alto, CA 94304}
\affil[c]{NASA Goddard Space Flight Center, 8800 Greenbelt Rd, Greenbelt, MD 20771}
\affil[d]{Space Telescope Science Institute, 3700 San Martin Dr, Baltimore, MD 21218}
\affil[e]{NASA Ames Research Center, Space Science \& Astrophysics Branch, Moffett Field, Mountain View, CA 94035}

\authorinfo{Send correspondence to Axel Potier: \href{mailto:axel.q.potier@jpl.nasa.gov}{axel.q.potier@jpl.nasa.gov}\\ © 2021. All rights reserved. }

\begin{document} 
\maketitle

\begin{abstract}
One of the primary science goals of the Large UV/Optical/Infrared Surveyor (LUVOIR) mission concept is to detect and characterize Earth-like exoplanets orbiting nearby stars with direct imaging. The success of its coronagraph instrument ECLIPS (Extreme Coronagraph for Living Planetary Systems) depends on the ability to stabilize the wavefront from a large segmented mirror such that optical path differences are limited to tens of picometers RMS during an exposure time of a few hours. In order to relax the constraints on the mechanical stability, ECLIPS will be equipped with a wavefront sensing and control (WS\&C) architecture to correct wavefront errors up to temporal frequencies $\gtrsim$1 Hz. These errors may be dominated by spacecraft structural dynamics exciting vibrations at the segmented primary mirror. In this work, we present detailed simulations of the WS\&C system within the ECLIPS instrument and the resulting contrast performance. This study assumes wavefront aberrations based on a finite element model of a simulated telescope with spacecraft structural dynamics. Wavefront residuals are then computed according to a model of the adaptive optics system that includes numerical propagation to simulate a realistic wavefront sensor and an analytical model of the temporal performance. An end-to-end numerical propagation model of ECLIPS is then used to estimate the residual starlight intensity distribution at the science detector. We show that the contrast performance depends strongly on the target star magnitude and the spatio-temporal distribution of wavefront errors from the telescope. In cases with significant vibration, we advocate for the use of laser metrology to mitigate high temporal frequency wavefront errors and increase the mission yield. 
\end{abstract}

% Include a list of keywords after the abstract 

\section{Introduction}
Since the first discovery of an exoplanet orbiting a Sun-like star in 1995, more than four thousand planets have been discovered through various methods. Direct imaging of these exoplanets remains an important technological challenge and has been limited so far to the detection of warm and massive bodies located at large distances from their stars ($>$10 AU). However, a primary goal of future space telescope missions will be to explore the habitable zone of extrasolar systems to detect the presence of Earth-like exoplanets (or ``exoEarths'') and determine the composition of their atmospheres via spectroscopy. Two of the four mission concepts currently under consideration by the NASA's 2020 Astrophysics Decadal Survey, the Large UV-Optical-InfraRed (LUVOIR) surveyor\cite{LUVOIR_finalReport} and the Habitable Exoplanet Observatory (HabEx)\cite{HabEx_finalReport}, seek to achieve this goal. LUVOIR is a large aperture space observatory with two main architectures: LUVOIR-A and LUVOIR-B, which are respectively 15-meter on-axis and 8-meter off-axis segmented telescopes. Both architectures are equipped with a coronagraph instrument called ECLIPS (Extreme Coronagraph for Living Planetary Systems). ECLIPS is designed to revolutionize the field of comparative planetary science by observing the nearest environment of $\sim$260 stars with a wide variety of spectral types and detecting $\sim$50 exoEarths. 

NASA's Exoplanet Exploration Program is currently leading the Segmented Coronagraph Design and Analysis (SCDA) study to assess the capability of segmented space-based telescopes in terms of exoEarth detection and characterization in the visible regime. The group has optimized coronagraph mask designs for segmented telescopes, demonstrated their abilities to reach contrast levels below $10^{-10}$ in simulation \cite{Juanola2019,Ruane2017_SPIE,Sirbu2020}, and predicted their scientific yields\cite{Stark2019}. As part of this work, the SCDA team also studied the sensitivity of these coronagraphs to optical aberrations and found that the contrast level of $\sim$10$^{-10}$ is maintained if the wavefront aberrations caused by segment phasing errors does not exceed $\sim$10~pm~RMS. Building on these past modeling efforts, we aim to assess the capability of a fast wavefront sensing and control (WS\&C) architecture, inspired by adaptive optics (AO) on ground-based telescopes, to correct dynamic aberrations using the combination of a deformable mirror (DM) and wavefront sensor (WFS), potentially relaxing the demanding constraints on the mechanical stability of the telescope. 

In Section~\ref{sec:Statistics_Wavefront}, we introduce the wavefront time series assumed for this study which represent dynamical aberrations based on a finite element model of a large segmented telescope. We project the resulting wavefronts into a modal basis for the purpose of adaptive optics correction. In Section~\ref{sec:AOsystem}, we develop an analytical model of an AO system and its temporal response. In Section~\ref{sec:OptimizationAndResults}, we optimize the response of such an AO system to correct for the wavefront aberrations and we derive the contrast performance of the instrument with respect to the stellar magnitude.

\section{Statistics of the dynamical wavefront}
\label{sec:Statistics_Wavefront}
\subsection{Finite element model of LUVOIR-A-like architecture}
To simulate dynamical aberrations at frequencies above 1~Hz, we analyze three different time series of the two-dimensional optical path difference (OPD) at the entrance pupil of the coronagraph instrument produced by Lockheed Martin (LM) that uses a finite element model of a large, segmented telescope representative of the LUVOIR-A architecture\cite{Dewell2019}. Each data set corresponds to $T=20$s of observation and is sampled with 8000 OPD maps resulting in an effective sample rate of 400~Hz. The OPD maps account for aberrations resulting from rigid body motion of the primary mirror segments and subsequent optics relative to each other.  As modeled, these perturbations are  driven by dynamic interaction of flexible structures with the noise and disturbances of the multi-stage pointing control system.

To reduce wavefront errors well below 1~nm rms, the model architecture includes a passive isolation at Control Moment Gyroscope (CMG) actuator mount and mechanical coupling reductions as well as a pointing stage that is designed to physically separate a sensitive science payload from its supporting spacecraft to reduce line of sight (LoS) errors. Pointing errors are sensed by the fine guidance mode of LUVOIR's High Definition Imager and controlled at two different rates by both the fast steering mirror and the Vibration Isolation and Precision Pointing System (VIPPS). VIPPS is a non-contact spacecraft-to-payload architecture wehereby payload inertial rigid-body attitude is controlled by voice-coil actuators, and payload-spacecraft relative alignment is sensed by non-contact sensors for real-time control\cite{Pedreiro2003}. The derived LoS error is lower than 0.1 milli-arcseconds, which is three times lower than the requirement specified in the LM SLSTD Phase 1 Final Report\cite{LMC2019}. While the LoS error is dominant in the overall error budget if uncontrolled, we assume that they are negligible in the following. Regarding the dynamic wavefront instability, CMG and VIPPS are assumed to introduce the same amount of aberration RMS and are the most dominant sources of noise. 

Three different time series were generated with RMS OPD error of 3.2~pm, 9.9~pm and 114~pm rms, which we refer to as sample A, B and C, respectively. Sample A represents a stable and optimistic scenario, sample B has roughly the level of error specified in the LUVOIR report\cite{LUVOIR_finalReport}, while sample C represents a more pessimistic case with 10$\times$ larger errors than B. The feasibility of these levels of stabilization with a 15-m segmented space telescope is beyond the scope of this paper. However, the telescope structure assumed in our model has not been anchored to test data or optimized for stability. Most importantly, metrology-based segment wavefront sensing and control models have yet to be incorporated into the integrated model. Therefore, some wavefront patterns contained in the data sets and later presented in this publication may be mitigated through a future modification of the mechanical design or via telescope metrology\cite{Lou2018}. For the purposes of this study, we ignore dynamic aberrations at temporal frequencies $\lesssim$1~Hz.

%In addition to high temporal frequency perturbations, quasi-static aberrations might also be an important limitation. These thermally induced aberrations are being simulated through a study performed by L3 Harris and Ball Aerospace, where mechanical reaction from thermal variation is considered. We assume these variations much slower than 1Hz such that they can be decoupled from the LM time series. The analysis of these quasi-statics aberrations and their mitigation will be performed in the near future but is beyond the scope of this study.

\subsection{Modal decomposition of a temporal series}
In AO theory, it is common to consider a modal decomposition of the OPD maps and correct each mode independently. The full AO system will effectively have independent control loops for each mode running in parallel to maximize efficiency. First, we decompose each OPD map $\phi(x,y,t)$ of the time series into a modal basis:
\begin{equation}
\label{eq:modaldecomposition}
\phi(x,y,t)=\sum_{i} a_i(t)Z_i(x,y)
\end{equation}
where $Z_i(x,y)$ represents the $i^{th}$ spatial mode of a chosen basis while $a_i(t)$ is its coefficient at a given time. %The above equation is always true for any chosen basis. But, for statistical reasons, it 
It is convenient to define the OPD map in an orthonormal basis where, for all modes $i,j$:
\begin{equation}
\label{eq:orthonormalbasis}
\frac{1}{\iint_\text{pupil}\,dx\,dy}\iint_\text{pupil} Z_i(x,y)Z_j(x,y)\,dx\,dy=\delta_{i,j}
\end{equation}
where $\delta$ is the Kronecker symbol:
\begin{equation}
\delta_{i,j}=\left\{
    \begin{array}{ll}
        1 & \mbox{if} \quad i=j \\
        0 & \mbox{if} \quad i \ne j
    \end{array}
\right..
\end{equation}
According to Eq. \ref{eq:modaldecomposition} and \ref{eq:orthonormalbasis}, the modal coefficients can be written as:
\begin{equation}
a_i=\frac{1}{\iint_\text{pupil}\,dx\,dy}\iint_\text{pupil} \phi(x,y,t)Z_i(x,y)\,dx\,dy.
\end{equation}
To calculate the spatial variance of the OPD maps at a given time:
\begin{equation}
\sigma^2(t)=\frac{1}{\iint_\text{pupil}dxdy}\iint_\text{pupil} \phi^2(x,y)\,dx\,dy=\sum_i a_i^2(t).
\end{equation}
The mean spatial variance over the time series is also easily calculated with an orthonormal basis. Indeed, if $E$ stands for the temporal ``expected value", we can write
\begin{equation}
\label{eq:temporal_spatial_variance}
\begin{aligned}
\sigma^2_\phi&=E\big[\sigma^2(t)\big]\\
&=E\left[\sum_i a_i^2(t)\right]\\
&=\sum_i E\big[a_i^2(t)\big]\\
&=\sum_i \sigma_{T,i}^2
\end{aligned}
\end{equation}
because the problem is linear. The last equation signifies that the averaged spatial variance of the OPD maps $\sigma^2_\phi$ can be calculated as the sum of the temporal variances of the modal coefficients $\sigma_{T,i}^2$. The $a_i$'s temporal variance can be calculated as the integral of the $a_i$'s temporal PSD. In the Section~\ref{sec:AOsystem}, we explain how an AO system acts on the temporal PSD of an incoming wavefront.

\subsection{Basis selection: Principal component analysis}
%AO systems equipping ground-based facilities often use either a Zernike basis to correct for the low order modes\cite{Jovanovic2015}, a Karhunen-Lo\`{e}ve\cite{Petit2014} (KL), or a Fourier\cite{Poyneer2016} decomposition for the high order spatial modes. Also, the emergence of large segmented telescopes could lead to use a basis where local Zernikes are decomposed on individual segments. In the case of LUVOIR, a Zernike decomposition over the entire pupil would not be relevant because Zernike polynomial's purpose is to translate the spatial statistics of the atmospheric turbulence. Moreover, we know that LUVOIR coronagraph performances are designed not to be affected by spatially low order modes.

%Since LM provided three temporal sets of OPD maps to analyze, we decide to develop a Principal Component Analysis (PCA) of the samples, which is an analogue to a discrete KL transform for atmospheric perturbations. 
We use principal component analysis (PCA) of the three time series individually to create an orthonormal basis which maximizes the variance of the phase projected on the first modes\cite{Roddier1999}. In that respect, PCA minimizes the mean square error that would result from a mode truncation. Numerically, the PCA is calculated using singular value decomposition (SVD) of the discrete OPD data set. If $X$ stands for the matrix representing the OPD data set, the SVD of $X$ can be written as:
\begin{equation}
X=U\Sigma V^T,
\end{equation}
where $\Sigma$ is a rectangular diagonal matrix with the singular values of $X$ sorted in ascending order. $U$ and $V$ are two square matrices whose vectors are orthogonal and called respectively left and right singular vectors. Under this formalism, the principal components $P$ are:
\begin{equation}
P=V^T.
\end{equation}

\begin{figure}[t]
    \centering
	\includegraphics[width=\linewidth]{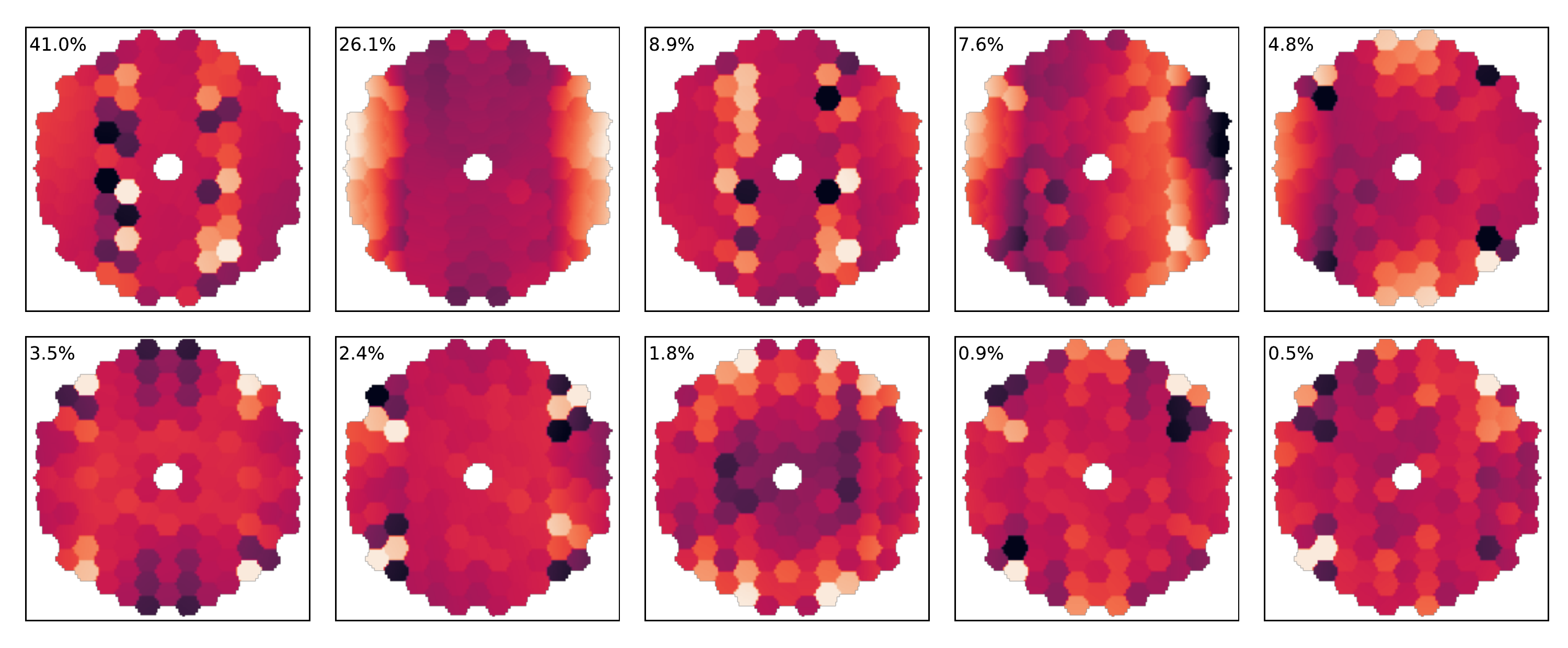}
	\caption{First principal components of the ``sample B" OPD time series ($\sim$10~pm~RMS). The number in the upper left of each panel is the percentage of the total variance represented by each mode.} % Figure caption
	\label{fig:PCA_modes} % Label for referencing with \ref{bear}
\end{figure}

Figure~\ref{fig:PCA_modes} shows the first ten principal components of the decomposition of time series B. 
%We apply PCA on the sample B provided by LM. The ten first principal components of this decomposition are shown in Fig.~\ref{fig:PCA_modes}. 
Here, the three first modes account for 76\% of the total variance and are characterized by a strong vertical shape; thus, the wavefront error stability is dominated by a small number of payload structural modes. These effects may be mitigated with a future mechanical optimization of LUVOIR A's architecture or by advanced telescope metrology. For instance, in our model, a 10x increase in damping of only 5 modes on the payload structure reduces the wavefront error to 2.5~pm. Since the eight first modes represent 96\% of the total variance, we will only control these eight modes in our analysis for sample B, which allows us to simplify subsequent numerical calculations with a minor loss of precision. Except for the second principal component, each mode is dominated by mid-spatial frequencies induced by local segment phasing errors. %This decomposition comfort the choice of not considering a low-order wavefront sensor where the major part of the variance would be missed. 

\begin{figure}[t]
    \centering
	\includegraphics[width=\linewidth]{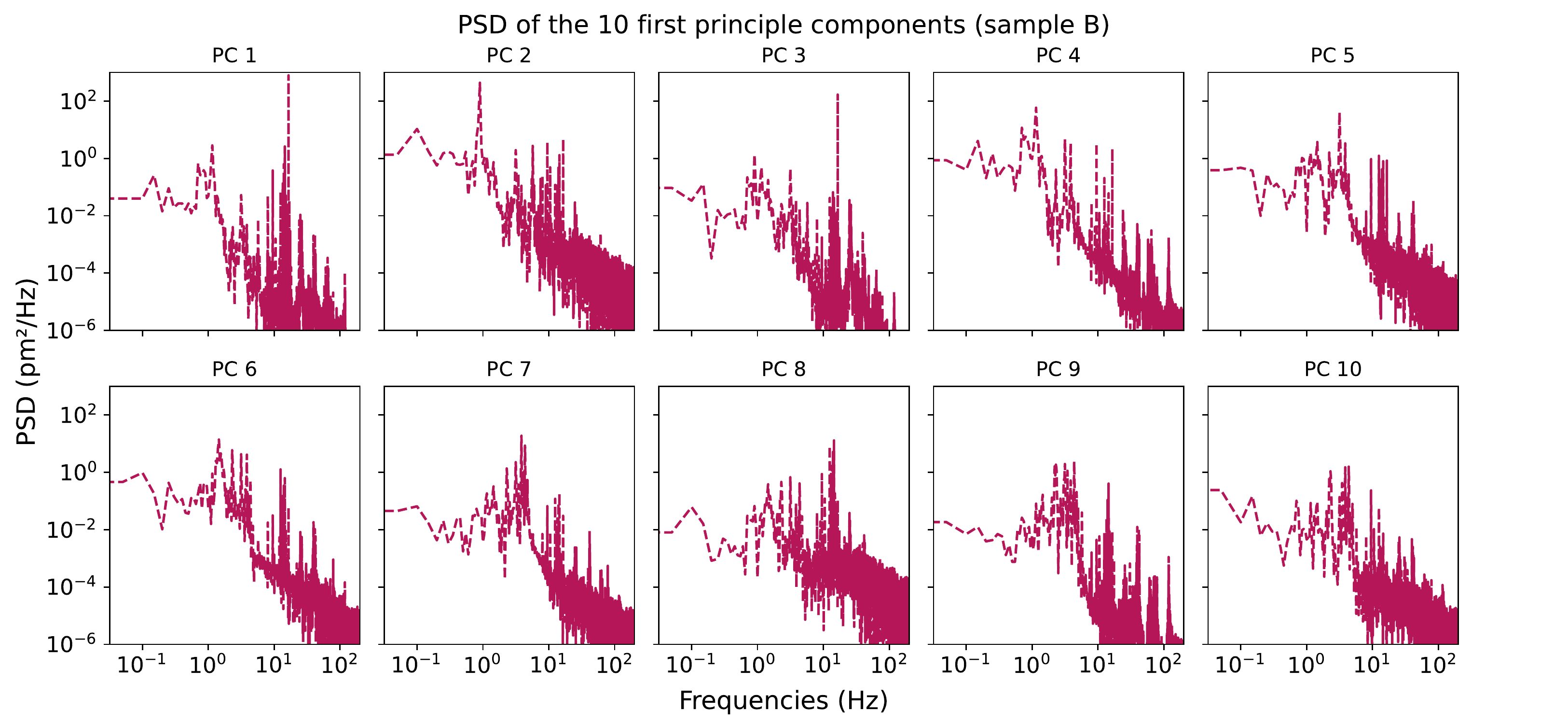}
	\caption{Power spectral densities of the 10 first principal components for the ``sample B" OPD time series.} % Figure caption
	\label{fig:PCA_PSD} % Label for referencing with \ref{bear}
\end{figure}

We calculate the temporal power spectral densities (PSDs) for each of the modes in the PCA basis. Figure~\ref{fig:PCA_PSD} shows the result for time series B. The PSDs obey power laws with exponents between -1.5 and -2 for the 10 first principal components. Moreover, the variances are dominated by a few localized vibrations at 16.5~Hz in the first mode and 0.9~Hz in the second. In fact, the 16.5~Hz vibration accounts for 20\% of the total variance. With this information in mind, there may considerable room for optimization of the telescope mechanical architecture to improve its dynamic stability by damping this, and other, specific vibration modes. Otherwise, the correction of these vibrations requires the AO system to be efficient at high temporal frequencies by increasing the AO loop bandwidth above 20~Hz.

%We have just presented some arguments advocating for a PCA in the context of fast Wavefront Sensing and Control (WS\&C). But there are also some drawbacks. 
There are some drawbacks to the basis we have selected. First, the AO system architecture is not taken into account when computing the PCA basis. In practice, a better option may be to adapt the basis such that the AO hardware optimally senses and corrects the first modes. For example, using our PCA approach, the DM may be unable to accurately correct the principal components because of its finite number of actuators and resulting fitting errors. Additionally, the WFS may not have the resolution to differentiate between the principal components accurately. However, we confirmed that the fitting error is negligible for a 64$\times$64 DM, as baselined for LUVOIR-A. We also optimize the number of pixels in the WFS to provide the best correction for the derived PCA modes. Secondly, correcting the first PCA modes in order of their OPD variance results in the maximum Strehl ratio, but not necessarily the best coronagraph contrast performance. Optimized coronagraphs for LUVOIR are, for instance, more sensitive to mid-spatial aberrations than to low order aberrations with equal variance. The PCA decomposition might not capture modes with low variance that have a high impact on the contrast performance of the instrument. In the following sub-section, we carry out contrast calculations using the PCA decomposition and confirm that the modes are mostly ordered by their impact of contrast despite these potential pitfalls. 

\subsection{Contrast calculation using principle components statistics}
We use an simplified end-to-end model of the ECLIPS coronagraph instrument to estimate the contrast change due to each OPD error map. The numerical simulation generates images for each of the 8000 OPD time steps with PROPER \cite{Krist2007} in a 20\% bandwidth around 550nm and using the optimized Apodized Pupil Lyot Coronagraph (APLC)\cite{Pueyo2019,Juanola2019}. For time series B, we find a mean intensity in the dark hole (DH) region, normalized by the maximum of the unocculted PSF, of $9.45\times10^{-11}$. Comparing this result with the normalized intensity induced by sample B that is reduced to its first 10 principal components, the mean difference between the normalized intensity at each time steps is $4.2\times10^{-13}$ and standard deviation of the differences is $8.8\times10^{-13}$. Considering only the first 10 principal components therefore induces $<$1\% error in the contrast calculation.

\begin{figure}[t]
    \centering
	\includegraphics[width=\linewidth]{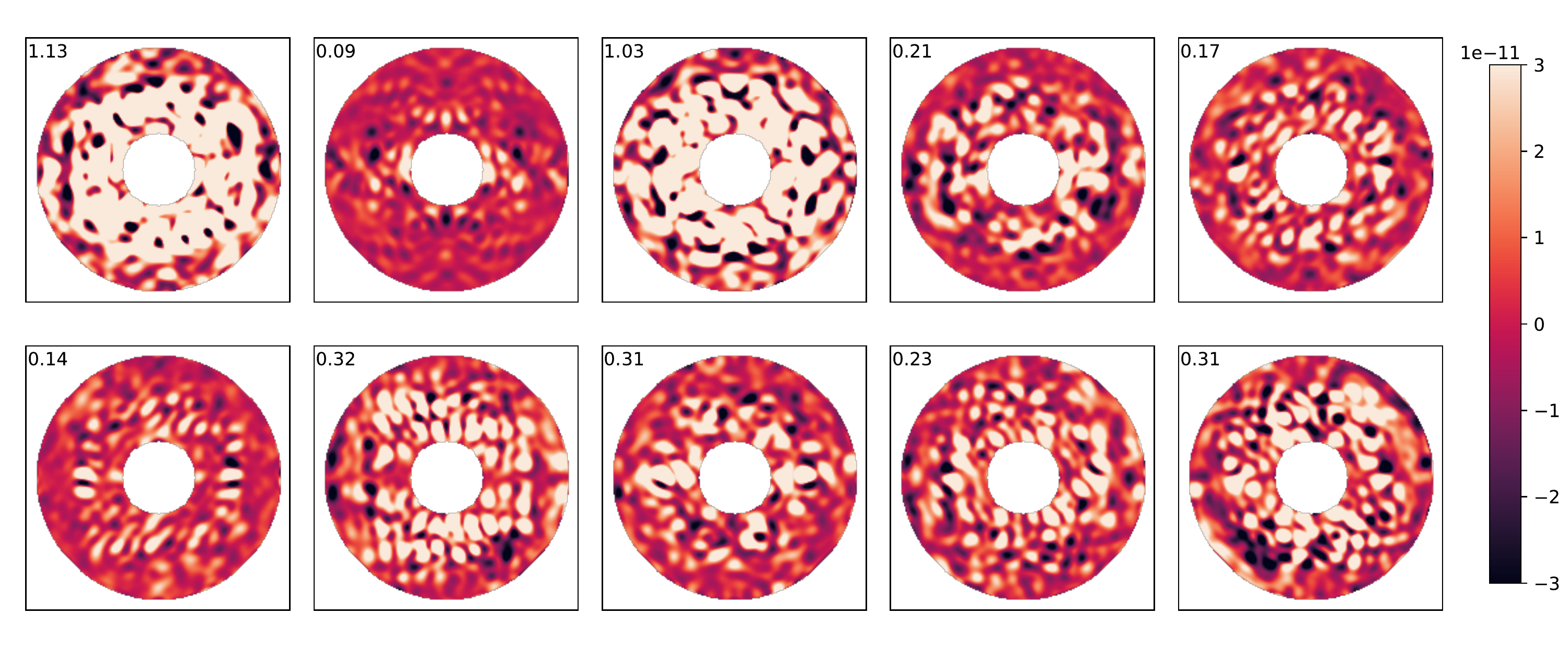}
	\caption{Contrast change in the image plane due to 10~pm rms of each principal component (ordered 1-10 from left to right). The normalized intensity calculated in the dark hole is also displayed in the upper left of each panel ($\times 10^{-10}$).} % Figure caption
	\label{fig:PCA_intensity} % Label for referencing with \ref{bear}
\end{figure}

\begin{figure}[t]
    \centering
	\includegraphics[width=8cm]{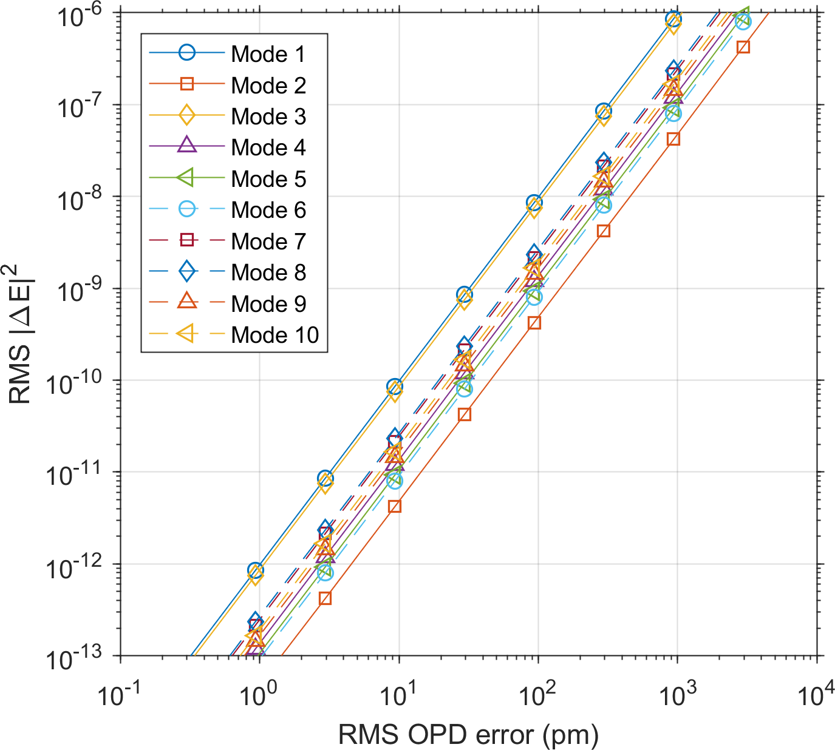}
	\caption{Normalized intensity induced by each of the first ten PCA modes with respect to the level of aberration.} % Figure caption
	\label{fig:DeltaE_vs_OPD} % Label for referencing with \ref{bear}
\end{figure}

In Fig.~\ref{fig:PCA_intensity}, we show the change in contrast induced by the first individual PCA modes when the spatial variance at coronagraph entrance pupil is equal to 10~pm rms. In Fig.~\ref{fig:DeltaE_vs_OPD}, we demonstrate the quadratic relationship between the normalized intensity $|\Delta E_i|^2$ induced by each individual mode $i$ at the science detector and the level of OPD error for the first 10 modes. This simple relationship allows us to easily estimate the contrast impact of the OPD data sets by scaling to any amount of OPD error. 

The contrast is most sensitive to the first and third principal components. Indeed, 10~pm rms of these modes individually bring the mean contrast in the DH above $10^{-10}$. %They have to be controlled carefully. 
Conversely, the second principal component has a relatively small effect on the contrast. % and thus a lower impact on the contrast error budget. 
In the ``sample B" time series, the order of magnitude of the contrast induced by the second component ($2.3\times10^{-12}$) is equivalent to the contrast induced by the fourth ($1.6\times10^{-12}$) despite the $>$3$\times$ larger OPD variance in the second mode. For the other components, the mean contrast values are almost identical at constant variance. This justifies using the PCA decomposition because the contrast impact of each mode in the time series (for the most part) decreases with the mode number.


The results above allow us to reduce the required computation time to estimate ECLIPS contrast performance. Indeed, we only need to pre-compute the $\Delta E_i$ due to each mode instead of numerically propagating the entire wavefront. At each time step, the modes are scaled by their coefficients $a_i(t)$ and the intensity at the detector is
\begin{equation}
    I(t) = \left|E_0 + \sum_{k=1}^{10} a_i(t) \Delta E_i \right|^2,
\end{equation}
where $E_0$ is the diffraction pattern of the coronagraph instrument without any OPD error. The variables $I$, $E_0$, and $\Delta E_i$ are functions of $x$, $y$, and $\lambda$. Since the frame rate of the science detector is much lower than $\sim$1~Hz, we are ultimately not interested in the instantaneous intensity. We are more interested in the time average spectral intensity representing long-exposure images and whose expression is given by: 
\begin{equation}
    I = \left\langle\left|E_0 + \sum_{i=1}^{10} a_i(t) \Delta E_i \right|^2\right\rangle.
\end{equation}
where $\langle.\rangle$ represent the temporal mean. This expression can be expanded as
\begin{equation}
\begin{split}
    I &= |E_0|^2\\ 
    &+ 2\sum_{i=1}^{10}\langle a_i\rangle \Re \{E_0^*\Delta E_i\}\\ 
    &+ \sum_{i=1}^{10}\sum_{i^\prime=1}^{10}\langle a_i a_{i^\prime}\rangle\Delta E_i \Delta E_{i^\prime},
\end{split}
\end{equation}
where the operator $\Re\{\}$ returns the real part of the field. 
To compute the long exposure images, we make assumptions about coefficient statistics. First, we assume each mode to be zero mean: $\langle a_i \rangle=0$. Also, we assume (and have confirmed by calculation) that the modes are independently distributed such that $\langle a_i a_{i^\prime}\rangle=0$ for $i \ne i^\prime$. Under these assumptions, the approximate long exposure image is calculated in the rest of this paper as:
\begin{equation}
\label{eq:MeanContrast_calculation}
    I = |E_0|^2\ + \sum_{i=1}^{10}\langle a_i^2 \rangle|\Delta E_i|^2, 
\end{equation}
where $E_0$ and $\Delta E_i$ are pre-computed while $\langle a_i^2 \rangle$ are equal to the temporal variances of each modes $\sigma^2_{T,i}$ in Eq.~\ref{eq:temporal_spatial_variance}. The stellar intensity at the science detector is therefore a linear combination of $|\Delta E_i|^2$ whose coefficients are the variance of each principal component. We seek to minimize this intensity by correcting the wavefront perturbations with a fast WS\&C system.

\section{Analytical response of the adaptive Optics system}
\label{sec:AOsystem}
\subsection{Wavefront Sensing and Control architecture}
In this context, the WS\&C system stabilizes the wavefront during science operations after the DH is created by minimizing both phase and amplitude static aberrations with conventional techniques\cite{GiveOn2007SPIE,GiveOn2011}. The fast wavefront sensor for this purpose would need to operate simultaneously with science observations, have sufficient spatial resolution, be sensitive to picometer level phase aberrations, and be in a common path configuration. One possible way to achieve this is to implement a Zernike Wavefront Sensor (ZWFS) built in to the coronagraph's focal plane mask (FPM), which reflects out-of-band light using a dichroic on the FPM substrate (while part of the in-band light can also be reflected by an occulting spot on the FPM) to the wavefront sensing detector. This in-situ ZWFS has been shown to be sensitive to picometer-level wavefront variations\cite{Ruane2020}. 

For the sake of simplicity, our simulation effectively assumes an ideal in-band ZWFS that collects all the light passing through the 20\% bandwidth visible filter. % and therefore leaves no signal for the science operation. 
In reality, the ZWFS would operate in a different band than science observation's filter, but this distinction is not relevant to our analysis. % but this assumption allows not accounting for any chromatic effects that might occur between the science and sensor optical paths. 
In accordance with the level of aberrations expected, we consider the relationship between measured intensity in the ZWFS and the phase to be perfectly linear during operations. 
For the correction, we ignore fitting errors due to the finite number of DM actuators and confirmed that the residuals are negligible for a 64$\times$64 DM and the OPD modes derived above. The in-pupil DM used to correct for the static aberrations is also used by the AO system to correct for the dynamical aberrations sensed with the ZWFS. The sensitivity of a ZWFS is often parameterized by $\beta$, which relates the phase error to the photon noise\cite{Guyon2005}. Here, we round $\beta$ to one at each wavelength and spatial frequency in a Fourier basis, while $\beta$ has recently been derived more precisely considering the realistic diffraction from the focal plane mask\cite{Ruane2020}. We generalize the calculation to a random orthonormal basis in section~\ref{subsec:WFSphotonnoise}. Our assumed telescope and WFS parameters are summarized in Table~\ref{table:TelescopeParameters}.

\begin{table}[htp]
	\caption{LUVOIR-A Specifications}
	\centering
	\begin{tabular}{lcc}
		\toprule
		Telescope diameter & $D$ & 15~m \\
		\midrule
		 Throughput & $\mathcal{T}$  & 0.3 \\
		 \midrule
		 Central wavelength & $\lambda$  & 545~nm \\
		 \midrule
		 Spectral bandwidth & $\Delta\lambda/\lambda$ & 20\% \\
		 \midrule
		 WFS sensitivity & $\beta$ & 1 \\
		 \midrule
		 WFS detector RON & $\sigma_{RON}$ & 1$e^-$ \\
		 \midrule
		 Vega flux (V band)& $\Phi_{Vega,ph}$ & 10$^8$ m$^{-2}$s$^{-1}$nm$^{-1}$\\
		\bottomrule
	\end{tabular}
	\label{table:TelescopeParameters}
\end{table}

\subsection{Transfer functions}
To stabilize the wavefront variation, a conventional linear AO system acts like a temporal filter on the modal coefficients $a_i(t)$. The corrected coefficients, $a^\prime_i(t)$, are given by the convolution between the impulse response and the input modal coefficient in the time domain:
\begin{equation}
\label{eq:input_output}
a^\prime_i(t)=h_i(t)\ast a_i(t),
\end{equation}
where $h_i(t)$ is the AO impulse response function, 
According to the convolution theorem, the Fourier transform of a convolution of two signals is the product of their Fourier transforms. Then, the square of the Fourier transform of Eq.~\ref{eq:input_output} gives the PSD:
\begin{equation}
\label{eq:TFdefinition}
PSD^\prime_i(f)=|H_i(f)|^2\cdot PSD_i(f),
\end{equation}
where $H_i(f)$ is the AO transfer function of the $i^{th}$ mode and $f$ is the temporal frequency. However, the AO system also introduces noise during the correction process. The resulting PSD of each mode is a function of the input wavefront error and noise:
\begin{equation}
\label{eq:PSDcorrection}
PSD^\prime_i(f)=|H_i(f)|^2\cdot PSD_i(f)+|H_N|^2\cdot N_i(f),
\end{equation}
where $N_i$ stands for the squared Fourier transform of the noise distribution introduced when sensing the $i^{th}$ mode. %Each of these terms need to be calculated to assess if the WS\&C architecture is able to help LUVOIR-A meeting its science specifications.

\subsubsection{Transfer function of each element}
\begin{figure}[t]
    \centering
	\includegraphics[width=0.85\linewidth]{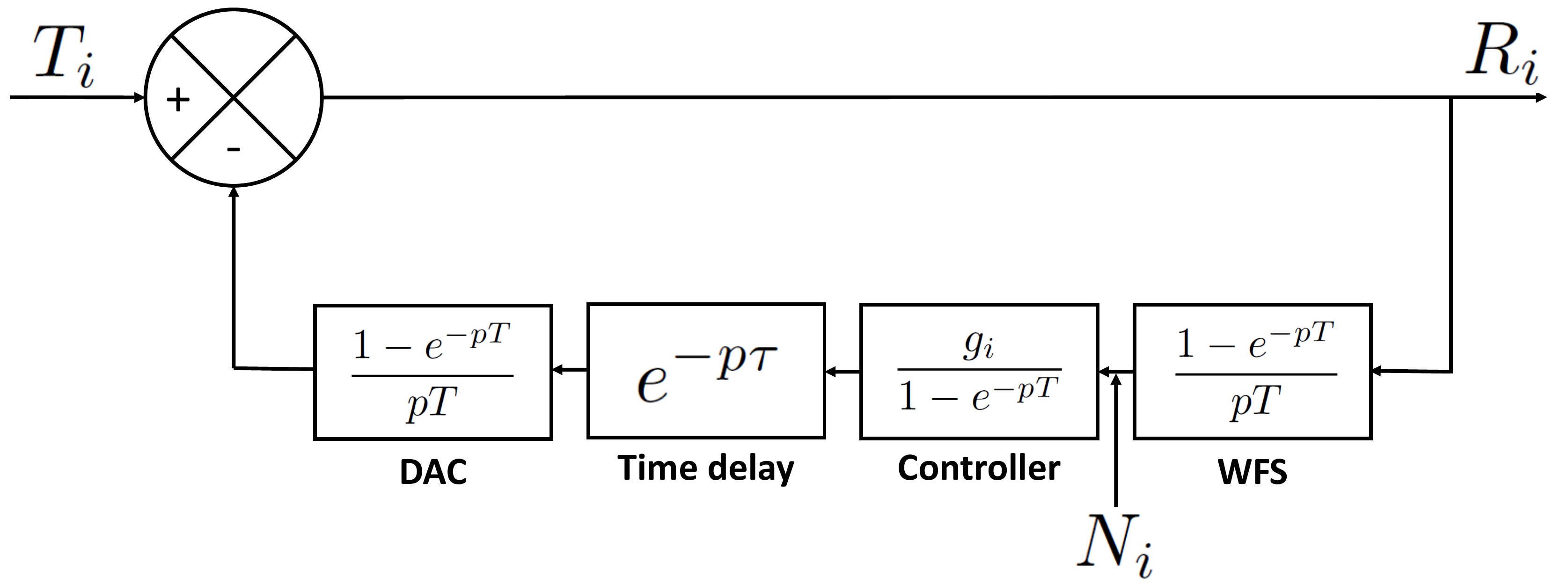}
	\caption{Block diagram of an AO control loop for a single mode in the case of a simple integrator.} % Figure caption
	\label{fig:AOloop} % Label for referencing with \ref{bear}
\end{figure}

The total rejection transfer function for one mode is the multiplication of the transfer functions of every serial element in the AO loop (see Fig.~\ref{fig:AOloop}). Here we consider only the wavefront sensor, the controller, the servo-lag error, and the digital-to-analog converter (DAC) which applies the signal to the DM. The DM transfer function is not taken into account here but could be considered for a future study. The transfer functions are calculated according to the methods outlined in the abundant literature from the ground-based AO community\cite{Gendron1994,Poyneer2016,Correia2017,Males2018b}. To express them, we use the Laplace formalism where $p=2i\pi f$. The Laplace transform properties are equivalent to the Fourier transform properties in this formalism.

\paragraph{Wavefront Sensor}
The wavefront sensor integrates the incoming perturbation for a period $T$ before passing along the information. Its impulse response is then a gate function of width $T$, corresponding to the servo loop period, centered at $T/2$. We can therefore write its transfer function as a sinc function multiplied by the Laplace transform of a time delay $e^{-i \pi f T}$:
\begin{equation}
H_{WFS}=\frac{1}{2}\text{sinc}(\pi f T)e^{-i \pi f T}=\frac{1-e^{-pT}}{pT}.
\end{equation}

\paragraph{Controller}
There are several options for the controller implementation. Here, we compare a simple integrator with a predictive control strategy as done in previous studies in different aberration regimes\cite{Males2018b}. We first consider a simple numerical integrator which is the conventional controller in AO. If $M_k$ is the $k^{th}$ WFS measurement, then the $k^{th}$ state for the $i^{th}$ mode, $C_{k,i}$, is
\begin{equation}
C_{k,i}=C_{k-1,i}+g_i M_{k,i}
\end{equation}
where $g_i$ is the servo-loop gain for the $i^{th}$ mode. Therefore, since $C_{k-1}$ is the state $C_k$ delayed by the servo-loop period $T$, we can write the relation between the integrator input and its output in Laplace coordinate:
\begin{equation}
C_i=C_ie^{-pT}+g_iM_i,
\end{equation}
or equivalently,
\begin{equation}
C_i=H_{con,i}M_i,
\end{equation}
where
\begin{equation}
H_{con,i}=\frac{g_i}{1-e^{-pT}}
\end{equation}
is the controller transfer function.

It has been demonstrated that the ideal rejection transfer function is proportional to the signal-to-noise ratio of the input wavefront as a function of frequency\cite{Dessenne1998}. A predictive controller shapes the AO transfer function in this way and aims to catch up with the global time delay introduced by the AO system, to improve the correction with respect to a simple integrator. This method is particularly efficient for correcting specific vibrations in the OPD time series. The algorithm is based on an autoregressive model where the estimated state $C_{k,i}$ depends linearly on its previous values as well as the previous measurements:
\begin{equation}
C_{k,i}=\sum_{n=1}^{q} b_{n,i} C_{k-n,i}+ \sum_{m=0}^{p} a_{m,i} M_{k-m,i}.
\end{equation}
This approach requires a good model of the wavefront perturbation entering the AO loop to minimize the residuals. Following the calculation described in the case of a simple integrator, we derive the temporal response of such a controller:
\begin{equation}
\label{eq:Predictive_controller}
H_{con,i}=\frac{\sum_{m=0}^{p} a_{m,i}e^{-mpT}}{1+\sum_{n=1}^{q}b_{n,i}e^{-nqT}}.
\end{equation}
This predictive controller reduces to the simple integrator when $p=0$, $q=1$ and $b_1=-1$.

\paragraph{Servo-lag error}
We have already shown the transfer function of a pure delay. For a delay $\tau$ introduced by numerical calculations and electronic read out time, the servo-lag transfer function is
\begin{equation}
H_\tau=e^{-p\tau}.
\end{equation}

\paragraph{Digital to Analog Converter}
The DAC is a zero-order holder whose role is to maintain a voltage on the DM until the voltages are updated by the controller (i.e. it holds the commands during the servo-loop period $T$). Thus, its impulse response is exactly equivalent to the WFS's impulse response and its transfer function can be written as
\begin{equation}
H_{DAC}=\frac{1}{2}\text{sinc}(\pi f T)e^{-\pi f T}=\frac{1-e^{-pT}}{pT}.
\end{equation}

\subsubsection{System frequency response}

The rejection transfer function $H_i$ links the AO loop residuals $a^\prime_i$ to the incoming perturbation $a_i$ (see Eq.~\ref{eq:input_output} and Eq.~\ref{eq:TFdefinition}). According to the scheme in Fig.~\ref{fig:AOloop}, the wavefront residuals are given by
\begin{equation}
R_i=T_i-H_{WFS} H_{con,i} H_\tau H_{DAC} R_i
\end{equation}
where $R_i$ and $T_i$ are the Fourier transforms of the corrected (residual) and input wavefronts, respectively. This equation can be rewritten as
\begin{equation}
\label{eq:correction_TF}
H_i=\frac{1}{1+H_{WFS} H_{con,i} H_\tau H_{DAC}}
\end{equation}
which is the transfer function of the wavefront correction system.

\begin{figure}[t]
    \centering
	\includegraphics[width=0.8\linewidth]{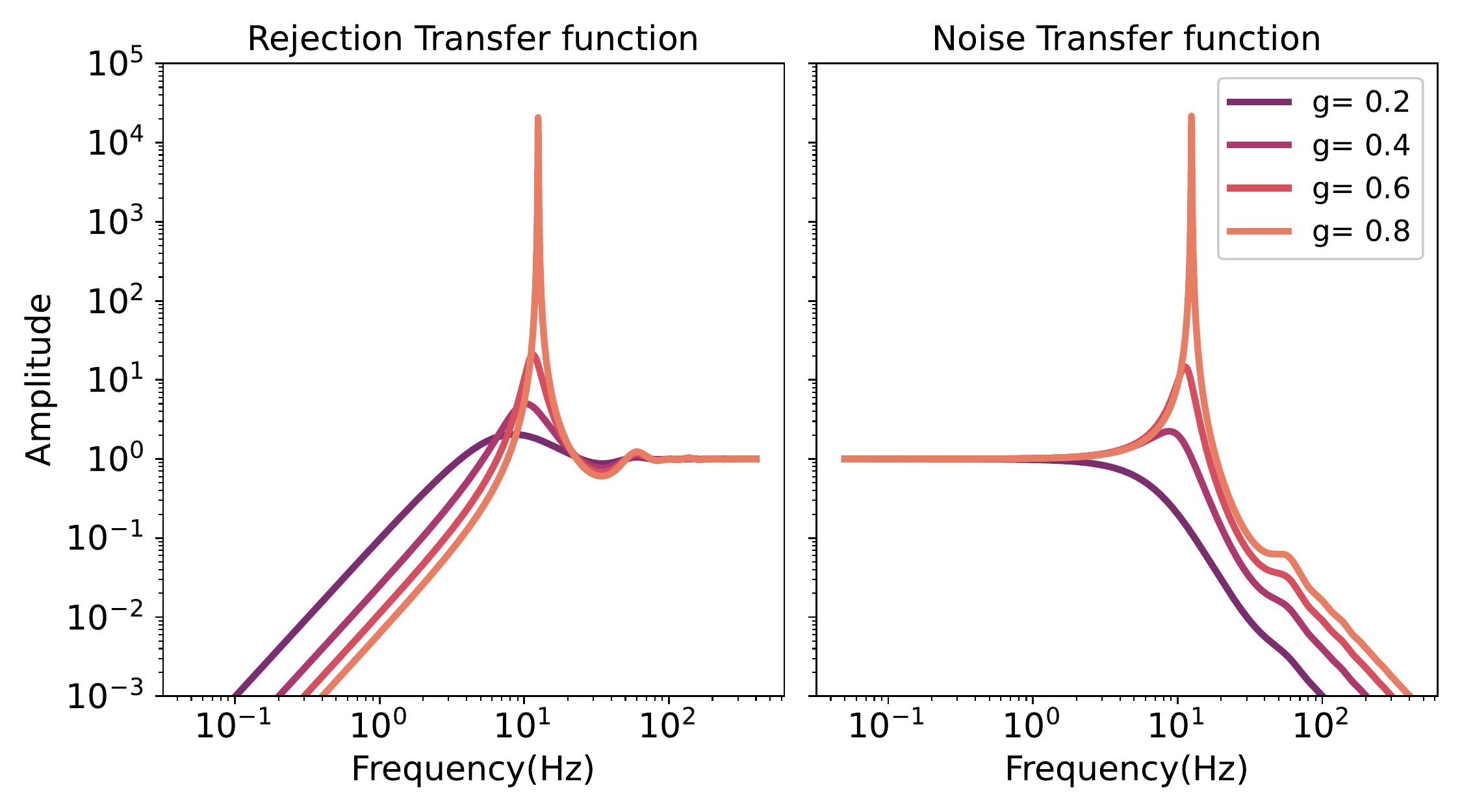}
	\caption{Squared modulus of the rejection transfer function (left panel) and noise transfer function (right panel) with varying AO gain. The sampling frequency is 100~Hz and the pure delay is equal to 1.5 frames.} % Figure caption
	\label{fig:TransferFunction} % Label for referencing with \ref{bear}
\end{figure}

We also derive an analytic expression for the noise transfer function only considering wavefront sensor noise (i.e. electronic read out noise) and photon noise injected before the controller as shown in Fig.~\ref{fig:AOloop}. The noise transfer function relates the residual signal introduced by the system in the presence of only noise. In that case, 
\begin{equation}
R_i= - H_{con,i} H_\tau H_{DAC} ( N_i + H_{WFS} R_i )
\end{equation}
and noise transfer function is
\begin{equation}
\label{eq:noise_TF}
H_N= - \frac{H_{con,i} H_\tau H_{DAC}}{1+H_{WFS} H_{con,i} H_\tau H_{DAC}}.
\end{equation}
Figure~\ref{fig:TransferFunction} shows the squared modulus of both the rejection and noise transfer functions with respect to the gain $g_i$ in the case of a simple integrator assuming a sampling frequency of 100~Hz and a pure delay of 1.5 frames. The integral below the rejection transfer function decreases with increasing the gain, while the integral below the noise transfer function increases with the gain. This means that the gain has to be optimized to balance the correction of the incoming wavefront without and the propagation of noise in the system. The gain can be optimized differently for each mode, depending on the modal PSD, in order to get the minimal residual:
\begin{equation}
\label{eq:optimize_SI}
g_i=\text{argmin}\,\sigma_{T,i}^2(g).
\end{equation}
The gain also defines the AO bandwidth. For instance, the AO bandwidth is below a tenth of the sampling frequency for a gain below 0.5. A signal faster than this frequency cannot be corrected. On the other hand, the signal is amplified for temporal frequencies whose order of magnitude is close to the sampling frequency, where the amplification depends on the gain. Thus, the gain optimization also aims to avoid any substantial amplification of the fast wavefront variation.

Predictive control can partially mitigate local vibrations in the perturbation's PSD that are faster than the AO bandwidth. However Bode's theorem imposes a strong restriction about the shape of the rejection transfer function \cite{Guesalaga2012}. Indeed, the integral below the rejection transfer function is fixed and only depends on the sampling frequency. It follows that the perturbation rejection tends to be worse in the vicinity of a vibration frequency with respect to the simple integrator case. The predictive controller also has to be shaped according to the input PSD. Good knowledge of the perturbation is therefore required to optimize the coefficient and minimize the residuals:
\begin{equation}
\label{eq:optimize_PC}
(a_0,...,a_p,b_1,...,b_q)_i=\text{argmin}\,\sigma_{T,i}^2(a_0,...,a_p,b_1,...,b_q)_i.
\end{equation}
This PSD knowledge requirement may limit the performance of predictive control on ground-based facilities. For example, observing conditions can quickly change subject to properties of the air turbulence in the atmosphere and telescope dome. As a result, the perturbation's PSD may therefore need to be evaluated too often for predictive control to be conveniently applied. Or, the spatio-temporal properties of the turbulence itself could limit the potential improvements provided by predictive control. On the other hand, space-based telescope offer much more stable and repeatable observing conditions and the coefficients may not require regular re-optimization. The wavefront errors may therefore be described by consistent and potentially well-characterized spatial modes and temporal PSDs.  

Both predictive control and simple integrator require knowledge of the perturbations to minimize the residuals. 
In practice, minimizing the integral in Eq.~\ref{eq:PSDcorrection} is equivalent to minimizing 
%It has been shown that minimizing the integral of Eq.~\ref{eq:PSDcorrection}, where $PSD_i$ represents the actual input perturbation, is strictly equivalent to minimize the criterion:
\begin{equation}
\mathcal{J}=\int |H_i(f)|^2\cdot PSD_{meas,i}(f) \, df,
\end{equation}
where $PSD_{meas,i}$ is the PSD measured by the AO wavefront sensor in open loop\cite{Dessenne1998}. The corresponding time series can therefore be measured directly and the parameters of the controller can be optimized based on those measurements. However, we use the full Eq.~\ref{eq:PSDcorrection} in this report since we already have prior knowledge of the OPD time series of aberrations impacting the telescope.

\subsection{Noise introduced by the wavefront sensor}
\label{subsec:WFSphotonnoise}
Before minimizing the integral of the residuals $PSD^\prime_i$ in Eq.~\ref{eq:PSDcorrection}, the shape and amplitude of the temporal noise introduced by the wavefront sensor $N_i$ needs to be simulated.

\subsubsection{Qualitative relationship between sensitivity and stellar flux}
\begin{figure}[t]
    \centering
	\includegraphics[width=5.6cm]{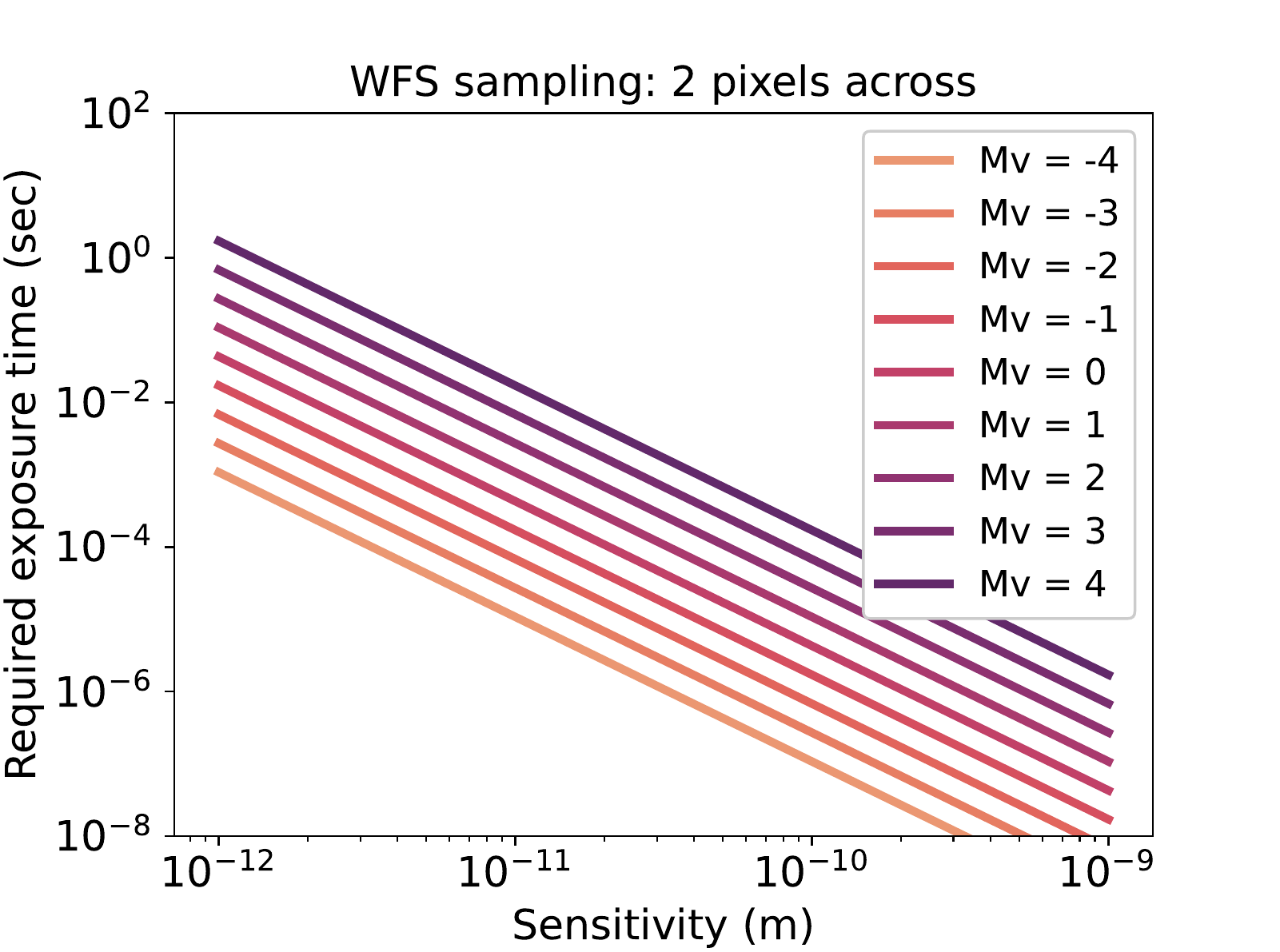}
	\includegraphics[width=5.6cm]{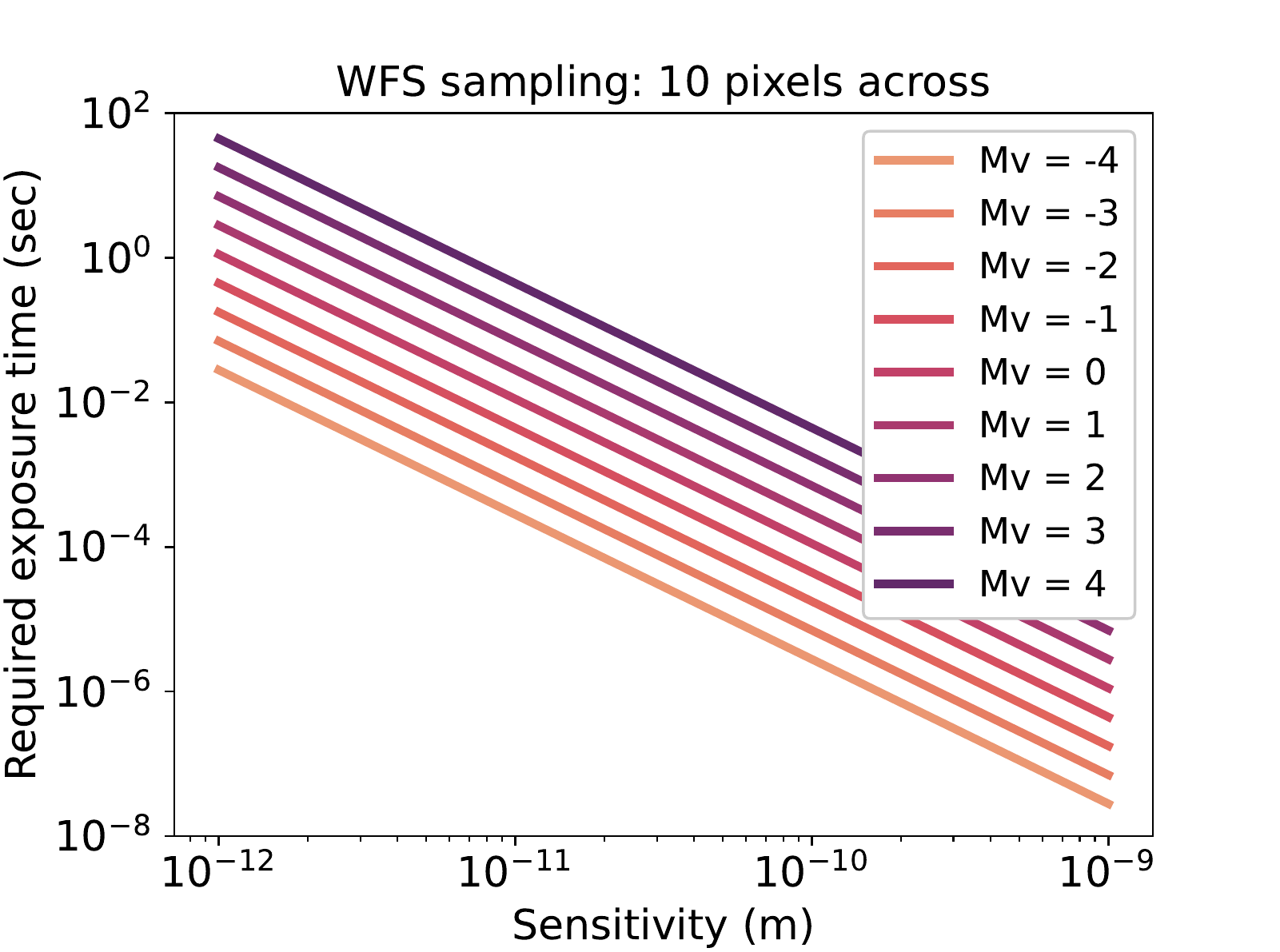}
	\includegraphics[width=5.6cm]{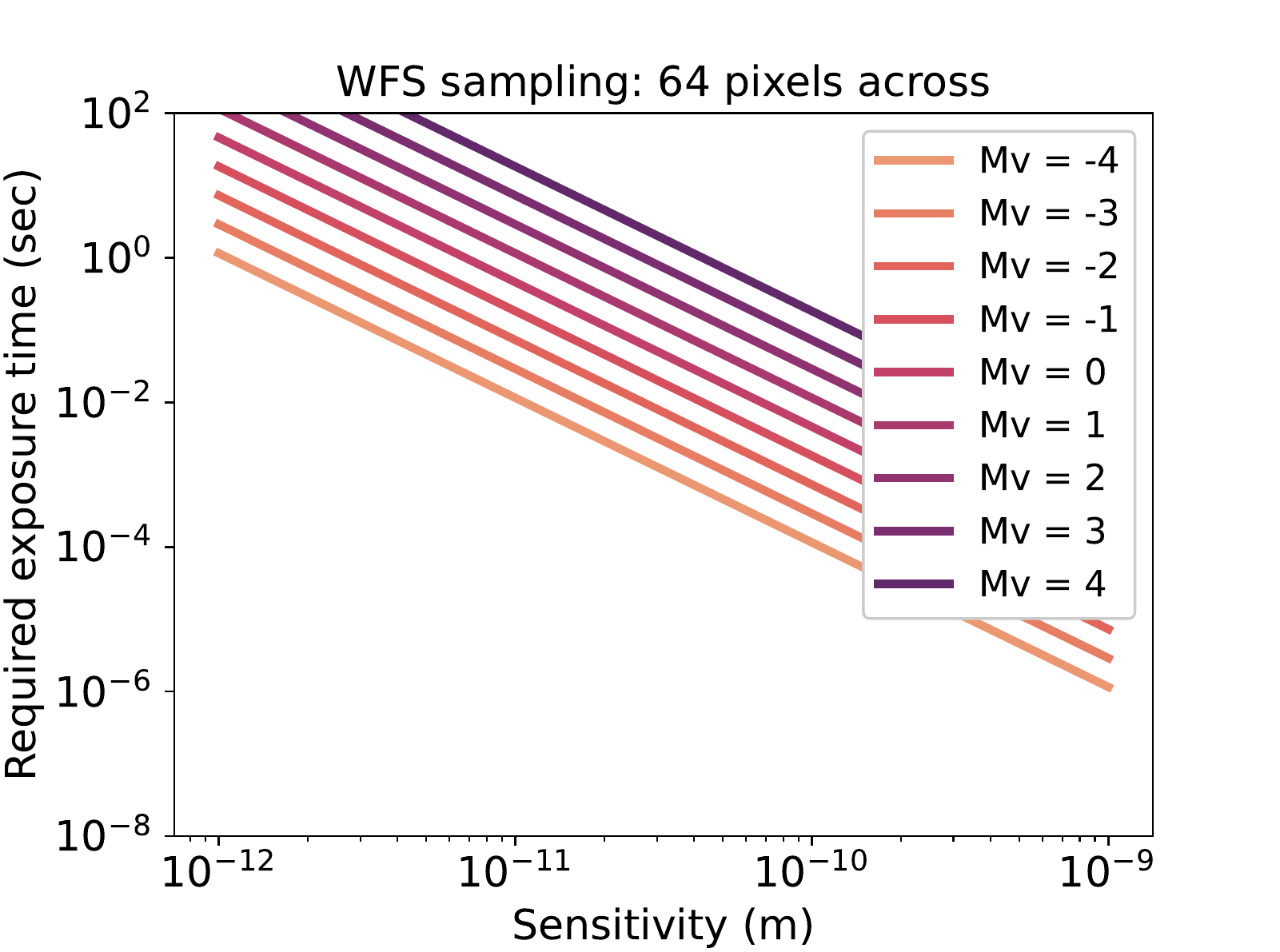} % Figure image
	\caption{Exposure time versus individual pixel sensitivity for a ZWFS. We consider 2 (left), 10 (center) and 64 (right) pixels across the pupil.} % Figure caption
	\label{fig:Exp_timevsSensitivity} % Label for referencing with \ref{bear}
\end{figure}

First, we understand qualitatively the impact of photon noise on the estimation with a ZWFS. For a convenient value of $\beta=\sqrt{2}$, the noise level on one pixel in radians is equal to $1/\sqrt{I_{pix}}$ where $I_{pix}$ represents the intensity collected on each pixel in units of photo-electrons. The sensitivity of the WFS then depends on the number of pixels across the pupil, the exposure time, and the magnitude of the observed star. Figure~\ref{fig:Exp_timevsSensitivity} shows the required exposure time of a visible WFS with respect to the desired sensitivity. For example, assuming a goal sensitivity of 1~pm per pixel, a Vega-like star would require an exposure time of 1~sec if the beam is sampled by 10 pixels. However, such a sampling limits the correction to low-order spatial modes. For a sampling of 64 pixels across, we see that almost 100~sec are required for an identical sensitivity and, thus, the correction would be limited to slower drifts and would leave faster wavefront variations uncorrected by the AO system. Another option would be to use a satellite equipped with a bright laser of negative magnitude to provide more photons\cite{Douglas2019}. Also, a lower sensitivity might be sufficient for the AO system to be effective and help increasing the exoplanet yield depending on the limiting wavefront errors. %To demonstrate it, the analytical modelling of the AO system in Eq.~\ref{eq:PSDcorrection} is required.

\subsubsection{Power spectral density of the detector noise}
%To model quantitatively the effect of the noise propagation in the AO system, we assume the detector to be affected only by photon noise and read out noise. 
Following previous literature\cite{Males2018b,Douglas2019}, we assume the noise on the WFS camera is uncorrelated both spatially and temporally since we only account for photon and read out noise. The detector noise can therefore be characterized as white noise with a flat PSD whose amplitude is defined by the variance per pixel $\sigma_n^2$. Even if the light distribution might not be flat on the WFS detector, the incoming noise is assumed identical on all pixels. The signal-to-noise ratio per pixel is therefore
\begin{equation}
S/N=\left(\frac{\sqrt{F_\gamma T+\sigma_{RON}^2}}{F_\gamma T}\right)^{-1},
\end{equation}
where $F_\gamma$ is the photon rate (photons/sec) collected by each WFS pixels and $\sigma_{RON}$ is the detector read-out noise. Since the Nyquist limit imposes the limits of integration on the PSD between 0 and $1/2T$, we write the temporal PSD measurement noise as
\begin{equation}
\int_0^{1/2T}N_N(f)\cdot df=(S/N)^{-2},
\end{equation}
where $N_N$ represents the PSD noise per pixel and finally:
\begin{equation}
\label{eq:Pixel_PSD_noise}
N_N=2(S/N)^{-2}T.
\end{equation}

\subsubsection{Propagation of noise through the system}
We simulate the propagation of the noise from the wavefront sensor following previous studies in the context of ground-based telescopes\cite{Rigaut1992,Gendron1994,Correia2020}. %If the WFS sensitivity $\beta$ is identical for all modes in the Fourier basis, this property is not verified in a more general case. The noise propagation is then different for each mode. To accommodate this, 
The WFS interaction matrix, $G$, of size $N_{pix}\times N_{modes}$ links the signal received on the WFS camera $s$ to the modal coefficient vector $a$ of the incoming phase:
\begin{equation}
\label{eq:WFS1}
s=G\cdot a.
\end{equation}
$G$ can be computed analytically thanks to an instrument model or it can be measured after the physical application of each mode in the pupil (with the DM, for instance). Then, in order to retrieve the modal coefficient of an incoming wavefront, Eq.~\ref{eq:WFS1} is inverted:
\begin{equation}
\label{eq:WFS2}
\tilde{a}=G^\dagger\cdot \eta,
\end{equation}
where $\tilde{a}$ stands for the estimate of $a$, $G^\dagger$ is the generalized inverse of $G$:
\begin{equation}
G^\dagger=(G^TG)^{-1}G^T,
\end{equation}
and $\eta$ is a vector of a pure noise on the WFS camera. The propagated noise covariance $\Sigma$ matrix is therefore defined as:
\begin{equation}
\label{eq:Covariance_matrix}
\begin{aligned}
\Sigma&=\tilde{a}\cdot\tilde{a}^T\\
&=G^\dagger \eta\,\eta^TG^{\dagger T}\\
&=(G^TG)^{-1}G^TG\left[(G^TG)^{-1}\right]^T\sigma_N^2\\
&=(G^TG)^{-1}\sigma_N^2.
\end{aligned}
\end{equation}
Finally, the noise PSD for the $i^{th}$ mode $N_i$ is the PSD of the noise reaching the WFS detector and then propagated through the WFS interaction matrix. According to Eq.~\ref{eq:Pixel_PSD_noise} and Eq.~\ref{eq:Covariance_matrix}, this PSD can be calculated as:
\begin{equation}
\label{eq:Mode_PSD_noise}
N_i = (GG^T)^{-1}\Big|_i \cdot 2(S/N)^{-2}T.
\end{equation}
The last equation means that the noise propagation coefficients for each mode are the diagonal terms of $(GG^T)^{-1}$. Since we need to invert the $G$ matrix, modes that are poorly estimated by the WFS result in drastically amplified noise. The number of pixels across the beam in the WFS detector is therefore an important parameter of this study. If the number of pixels in the WFS is too low, the noise propagation per mode increases. However, increasing the number of pixels decreases the signal-to-noise ratio on the detector. This trade off is studied in Sec.~\ref{subsec:Contrast_Performance}. Nevertheless, the optimal number of pixels across the beam actually depends on the chosen OPD decomposition. A low-order wavefront sensor with few pixels would be highly sensitive but may be blind to important aberrations, like segment phasing errors. More pixels would decrease the global sensitivity of the system but allows for correction of mid-spatial frequency perturbations such as those induced by segment phasing errors.

\begin{figure}[t]
    \centering
	\includegraphics[width=\linewidth]{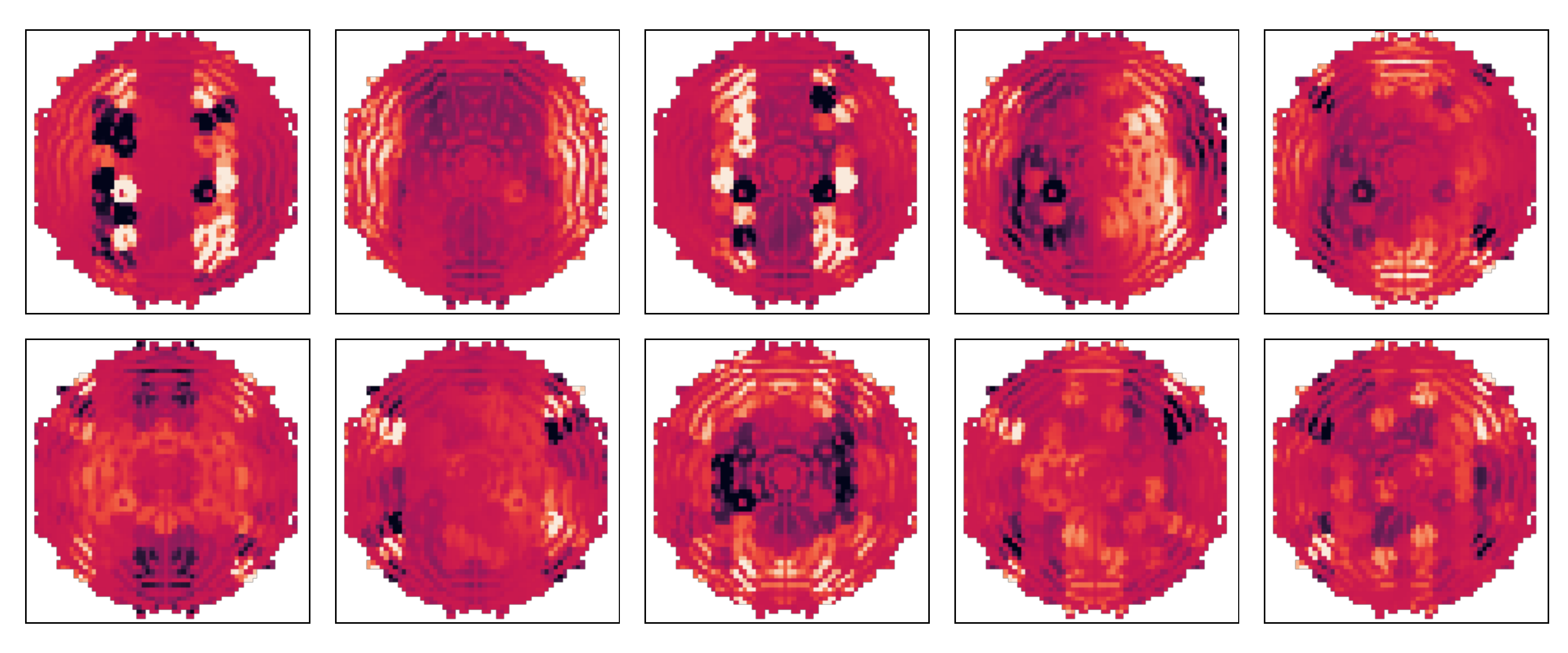}
	\caption{Change in intensity on the ZWFS detector due to each principal components of the OPD time series assuming 64 pixels across the beam.} % Figure caption
	\label{fig:WFS_images} % Label for referencing with \ref{bear}
\end{figure}

\begin{figure}[t]
    \centering
	\includegraphics[width=9cm]{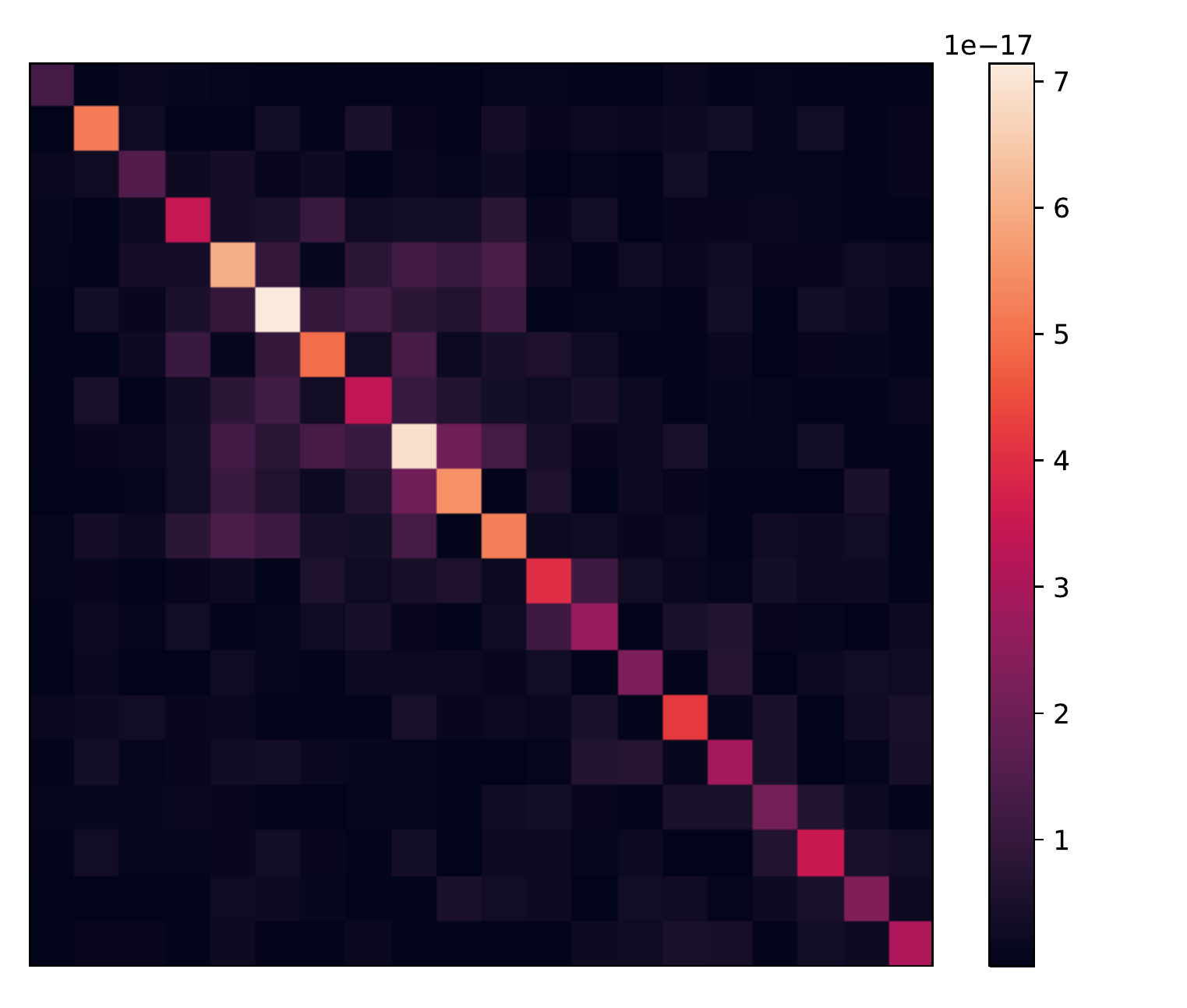}
	\caption{Covariance matrix of the twenty first principal components of sample B time series, sensed by a ZWFS. In the WFS detector plane, there are 64 pixels across the pupil. The first mode is in the upper left.} % Figure caption
	\label{fig:Covariance_Matrix} % Label for referencing with \ref{bear}
\end{figure}

We illustrate these principles by calculating the matrix $(GG^T)^{-1}$ in Eq.~\ref{eq:Covariance_matrix}, which is also known as the covariance matrix. For a WFS capable of sensing each PCA mode separately, this matrix should be diagonal. To generate it, we use PROPER to model a version of the LUVOIR-ECLIPS instrument equipped with a ZWFS and a WFS camera with 64 pixels across the pupil. The generated WFS images are shown in Fig~\ref{fig:WFS_images} and the covariance matrix is shown in Fig.~\ref{fig:Covariance_Matrix}. In this case, the covariance matrix is not perfectly diagonal and the modes are partially cross-correlated. It means that an individual component inject in the pupil may be confused with other modes. When looking at individual ZWFS images in Fig.~\ref{fig:WFS_images}, the PCA modes are qualitatively recognizable, but the undersampling of both the primary mirror and the shaped pupil apodizer has a non-negligible impact on the WFS. Again, this noise can be mitigated by increasing the WFS sampling, but that would result a lower SNR at each pixel for an equivalent stellar magnitude. In the following sections, we assume the covariance matrix to be diagonal and we keep the diagonal terms as generated in Fig.~\ref{fig:Covariance_Matrix}.

\section{Optimized performance of the adaptive optics system}
\label{sec:OptimizationAndResults}
\subsection{Residual variance and stellar magnitude}
\subsubsection{Optimized integrator}

\begin{figure}[t]
    \centering
	\includegraphics[width=8.5cm]{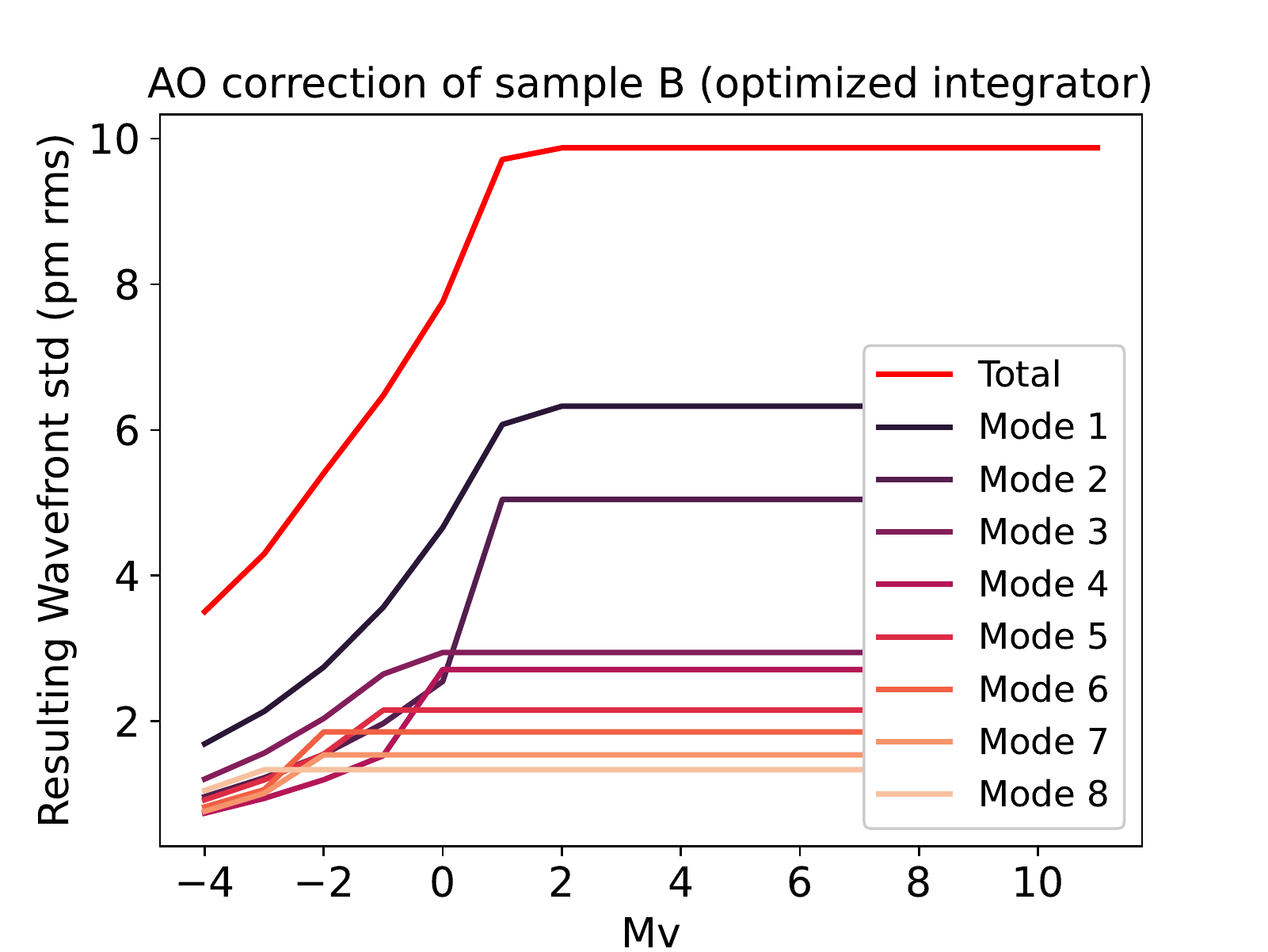}
	\includegraphics[width=8.5cm]{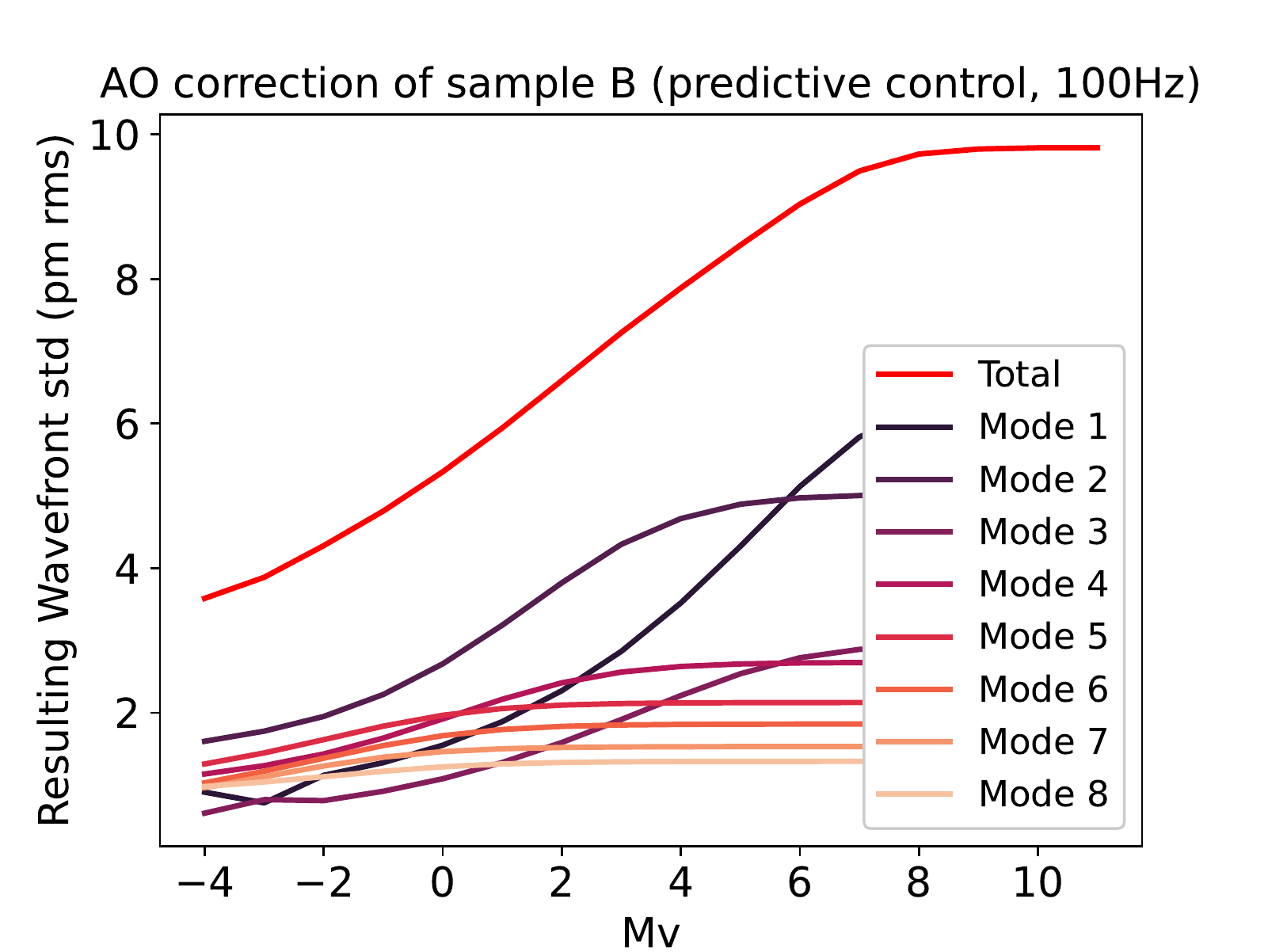}
	\caption{Left: Standard deviation of the residual wavefront after a conventionnal AO correction (optimized integrator) running at 1~kHz. Right: Standard deviation of the residual wavefront after a predictive control correction whose system is running at 100~Hz. The input dynamical aberrations is sample B. The residual of the eight first principal components is also shown.} % Figure caption
	\label{fig:Variance_Magnitude} % Label for referencing with \ref{bear}
\end{figure}

According to Eq.~\ref{eq:PSDcorrection} we now have all the required information to apply a correction on the modal PSDs derived in section \ref{sec:Statistics_Wavefront}. The PSD noise is calculated according to Eq.~\ref{eq:Mode_PSD_noise} while the AO transfer functions are calculated through Eq.~\ref{eq:correction_TF} and Eq.~\ref{eq:noise_TF}. We have shown that the success of a correction strongly depends on the photon noise and thus on the stellar magnitude. We therefore apply the filtering independently for stellar magnitudes between -4 and 11. First, we assume the controller is a simple integrator, the servo-loop frequency sampling is 1~kHz, and the WFS is sampled with 64 pixels across the beam. We also assume 1.5 frames of pure delay. The other relevant parameters are given in Table~\ref{table:TelescopeParameters}. The gain is optimized for each mode by applying a Broyden-Fletcher-Goldfarb-Shanno (BFGS) algorithm. For a wavefront error following the statistics of sample B, we plot the residual wavefront standard deviation as well as each mode residual standard deviation with respect to the stellar magnitude in Fig.~\ref{fig:Variance_Magnitude}. For an initial wavefront variation of about 10~pm rms, the AO system brings an improvement when the magnitude of the observed star is $<$1 (i.e. brighter than 1). Above this magnitude (i.e. fainter than 1), the modal noise PSD is too high for the AO system to correct any modes and all the modal gain are set to 0. Between $M_V=0.2$ and $M_V=1$, only the first principal component is controlled by the AO system as expected. Indeed, the ratio between $PSD_I$ and $N_i$ in Eq.~\ref{eq:PSDcorrection} is the highest for this mode as desired and built-in to the PCA decomposition. This means that, for such a pixel sampling and exposure time, no other mode or decomposition can be corrected for stellar magnitude above 1. A longer exposure time could be used to correct slow varying drifts when observing stars at higher magnitudes. However, most of the OPD variance would remain uncorrected and the AO system would not bring any significant modifications in the OPD statistics. This result is extremely important. It means a laser guide star is required to stabilize all the mid-order spatial modes with an AO system under 10~pm rms if the controller is an integrator. We expect the total standard deviation would divided by two for $M_V=-2.3$. 

\subsubsection{Predictive control}
\begin{figure}[t]
    \centering
	\includegraphics[width=\linewidth]{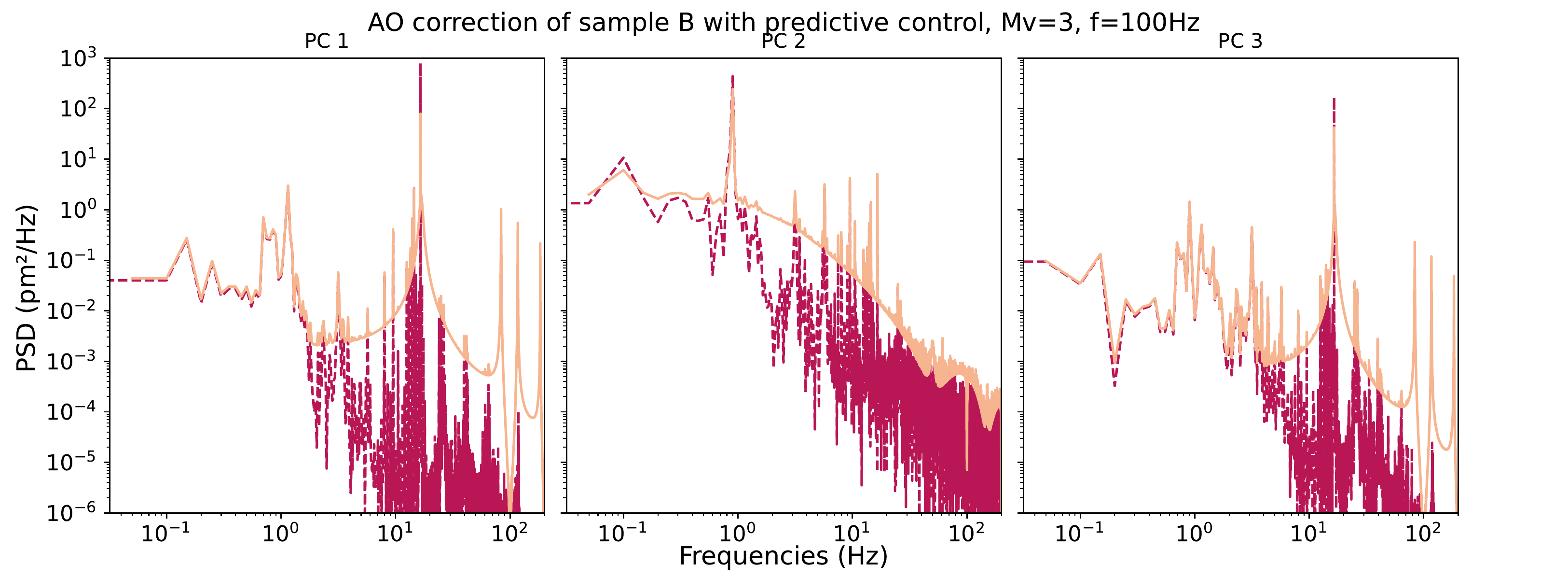}
	\caption{Power Spectral Densities of the three first principal components of sample B before (purple) and after (yellow) correction with an AO system running at 100Hz and equipped with a predictive controller. The wavefront sensor is sampled at 64 pixels across the beam and the magnitude of the target star is equal to 3 in V-band.} % Figure caption
	\label{fig:PCA_PSD_PC} % Label for referencing with \ref{bear}
\end{figure}

The numerical simulations above assumed an optimized integrator. We expect better performance with predictive control since it should be able to focus on mitigating vibrations (e.g. at 0.9~Hz and 16.5~Hz in sample B). We plot in Fig.~\ref{fig:PCA_PSD_PC} the PSD of the three first principal components before and after correction with predictive control for $M_V=3$. The loop is assumed running at 100~Hz in order to increase the signal-to-noise ratio on each pixel by $\sqrt{10}$ with respect to the simulation in Fig~\ref{fig:Variance_Magnitude}. We choose $p=q=2$ in Eq.~\ref{eq:Predictive_controller} which implies that we optimize five different parameters in Eq.~\ref{eq:optimize_PC}. We use a constrained minimization method to minimize the cost function in the frequency domain and the parameters are all bounded in between -1 and 1, except $a_0$ whose value is between 0 and 1. While these moderated bounds and algorithms have shown reliable results, the use of recursive algorithms applied in the time domain such as a LQG/Kalman estimators \cite{Leroux2004,Kulcsar2006,Poyneer2010} whose parameters are optimized by solving a Riccati's equation \cite{Guesalaga2012,Males2018b} is essential to ensure the stability and robustness of the correction. Their implementation is beyond the scope of this publication but might be important for a more detailed model of the performance and limitations of such a controller with LUVOIR.

We demonstrate that predictive control helps correcting one specific vibration for each mode in Fig.~\ref{fig:PCA_PSD_PC}. To make such correction at particular frequencies, the signal had to been amplified at other frequencies, as expected from Bode's theorem. However, the variance residuals has been minimized overall. The right panel of Fig.~\ref{fig:Variance_Magnitude} shows the standard deviation of the residuals with respect to the observed star magnitude. Predictive control is able to minimize the wavefront variance with stars fainter than $M_V=7$ which constitutes the majority of the natural stars targeted by LUVOIR. This represents an important improvement with respect to the optimized integrator.

\subsection{Contrast predictions}
\label{subsec:Contrast_Performance}
\begin{figure}[t]
    \centering
	\includegraphics[width=10cm]{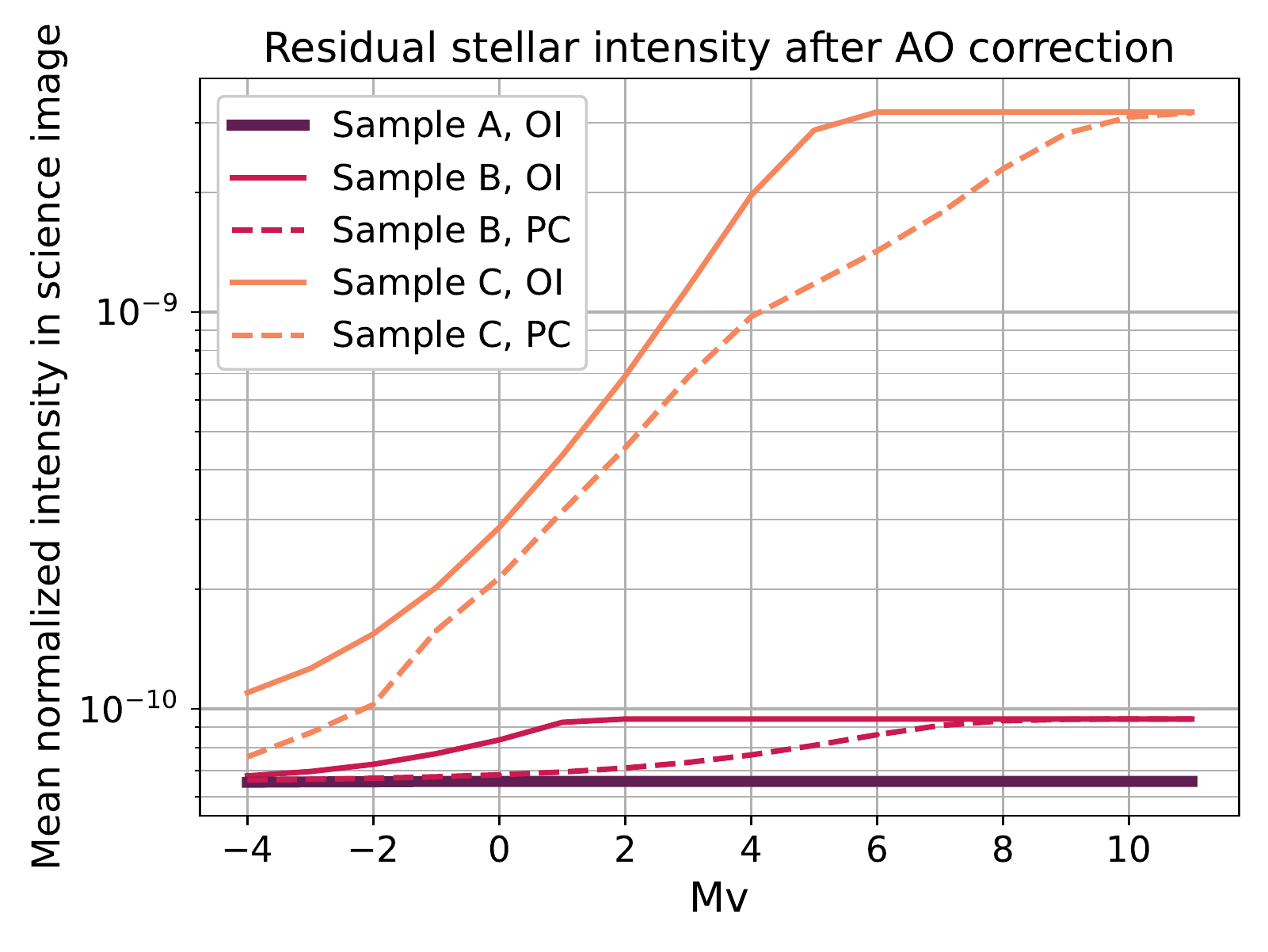}
	\caption{Mean normalized intensity in the dark hole after the wavefront being corrected by an AO system. The AO performance depends on the magnitude of the observed star and on the initial level of turbulence. Purple, pink and orange curves use respectively sample A, B and C as inputs of the simulations. OI stands for optimized integrator while PC means predictive control.} % Figure caption
	\label{fig:Contrast_vs_magnitude} % Label for referencing with \ref{bear}
\end{figure}

We reproduced an equivalent analysis for the A and C time series where we find a similar but not identical PCA decomposition. The modes are then corrected independently with optimized AO systems. For the three samples, Figure~\ref{fig:Contrast_vs_magnitude} shows the normalized intensity in the image plane of the APLC coronagraph, as calculated in Eq.~\ref{eq:MeanContrast_calculation}, after the AO system versus the magnitude of the observed star. Samples A, B and C's residuals are all performed with an optimized integrator running at 1~kHz. We also present the normalized intensity when the times series B and C are corrected with predictive control. The best contrast result between a servo-loop frequency of 100~HZ and 1~kHz is kept at each magnitude. 

First, the figure shows that adaptive optics is not required in the best case scenario (sample A). Indeed, the wavefront is already low enough before any correction for the normalized intensity to be dominated by the diffraction pattern of the APLC. Sample B's scenario is quite different. If the overall budget error of LUVOIR-A imposes a contrast below $10^{-10}$, minimizing dynamical aberrations is required because they already dominate the error budget since they brings the normalized intensity to $9.45\times10^{-11}$. We show that using a simple integrator would require a laser guide star to achieve a significant correction. A $M_V=-4$ star would reduce the contrast down to sample~A contrast performance. The coronagraph performance would highly benefit from predictive control to mitigate strong vibrations in the first and third principal components. For $M_V<8$, the variance of these two modes is reduced which brings about a major decrease in the dynamical aberrations in the error budget. Finally, in the worst case scenario where the wavefront is above 100~pm rms (sample C), the AO system is able to correct aberrations with natural guide stars, even with an optimized integrator. However, predictive control provides better contrast results. Again, there are two different regimes here. For natural guide stars, the AO loop should run at 100~Hz because it brings about better contrast results than the 1~kHz servo loop when the WFS is photon starved. But, this kind of correction remains limited to brighter stars since the pm level sensitivity has been already reached while high frequency vibrations remain uncorrected. Then, decreasing the WFS exposure time to correct for these vibrations when observing brighter stars improves the performance. With such an optimization and the essential contribution of a laser guide star, the normalized intensity would be capable of remaining lower than $10^{-10}$ in this worst case scenario. To avoid introducing additional spacecrafts for the laser guide star, future telescope design studies should therefore consider strategies that minimize the overall dynamical aberrations levels to a few dozens of pm rms and then take advantage of the internal AO system.

\subsection{Discussion}
\subsubsection{Optimization of the wavefront sensor spatial sampling}
\begin{figure}[t] 
    \centering
	\includegraphics[width=14cm]{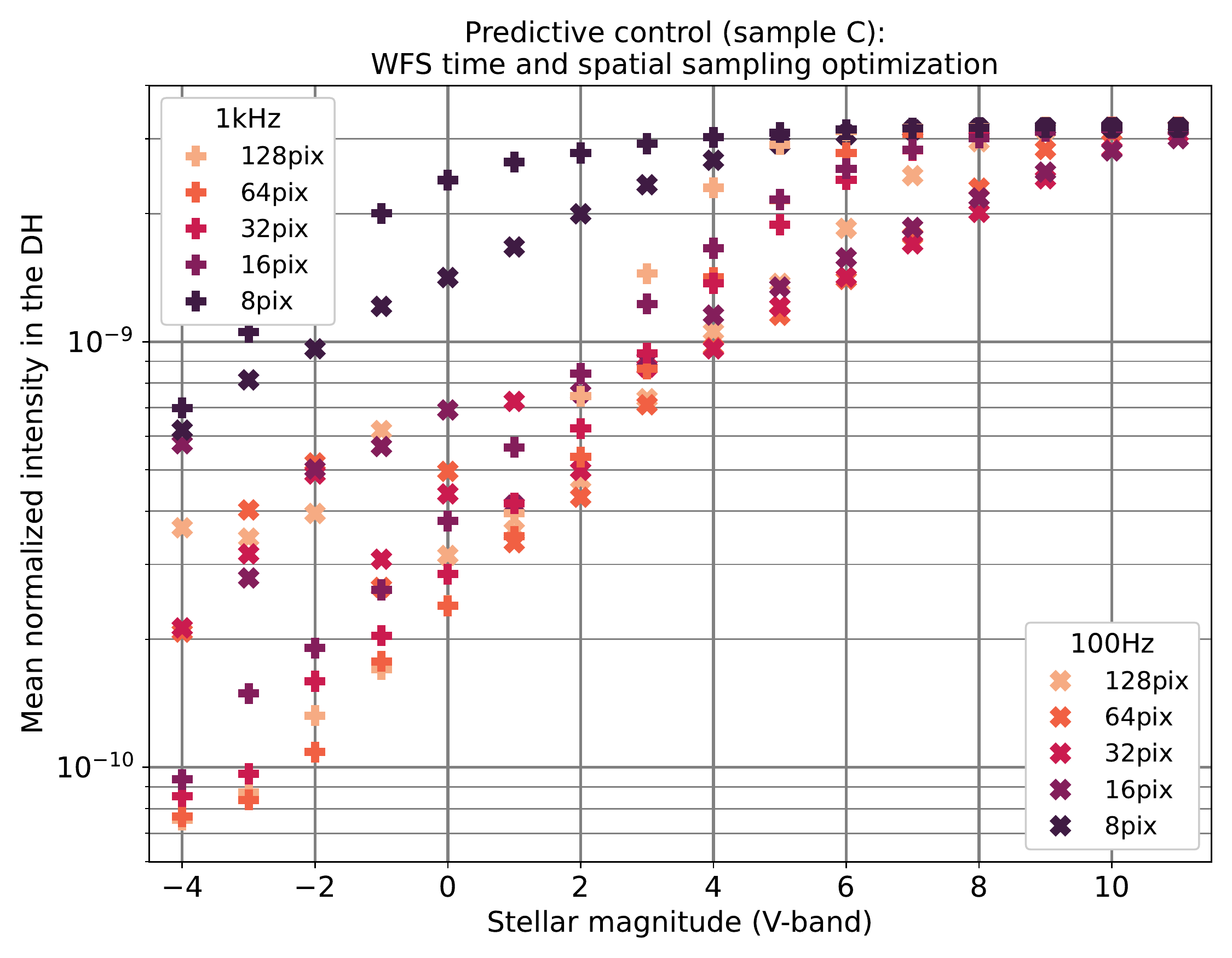}
	\caption{Normalized intensity versus stellar magnitude as a function of the adaptive optics system characteristics. Predictive control under sample C stability conditions is assumed.} % Figure caption
	\label{fig:Performance_vs_Npix} % Label for referencing with \ref{bear}
\end{figure} 
In Section~\ref{subsec:WFSphotonnoise}, we discussed the trade off between the spatial sampling of the WFS camera and the photon noise per pixel. We show here that the overall AO system can be optimized by modifying this sampling based on the observing conditions. Figure~\ref{fig:Performance_vs_Npix} shows the performance of the adaptive optics system with respect to the number of pixels across the beam in the WFS plane. We assume five scenarios where $N_{pix}=$8, 16, 32, 64 or 128 pixels across the beam. This could be interpreted as a WFS camera that can bin pixels based on the observing scenario. Doubling the amount of pixels divides the signal-to-noise ratio at each pixel by 4. Using 8 pixels across multiplies the signal-to-noise ratio by 64 with respect to the 64-pixel scenario. At equal sensitivity, it would represent a shift of 4.5 magnitudes of the curves in Fig.~\ref{fig:Contrast_vs_magnitude} toward the right. However, the sensitivity of the WFS decreases drastically since it estimates mid-spatial frequencies with fewer pixels; as a result, $(GG^T)^{-1}$ increases in Eq.~\ref{eq:Mode_PSD_noise} and so does the modal noise. Therefore, there is a trade off between signal-to-noise ratio and modal sensitivity. These parameters can be optimized with respect to the source magnitude. For natural guide stars with magnitude $M_V\geq4$, Fig~\ref{fig:Performance_vs_Npix} shows that a WFS running at 100~Hz and with 32~pixels across gives the best contrast performance. For instance, we reach a normalized intensity of $1.4\times10^{-9}$ with sample C initial perturbation and a stellar magnitude $M_V=6$. For natural guide star where $0<M_V\leq4$, better performance require to increase the number of pixels across the pupil. Finally, optimizing the servo loop would require us to use 64 pixels across the beam and an exposure time of $10^{-3}$s to obtain the best performance with a laser guide star ($M_V<0$). At $M_V=-4$, 128 pixels across would slightly improve the performance with respect to the 64 pixels case.

\begin{comment}
\subsubsection{Incidence of the servo-lag error}
All the results present above assume 1.5 frames of pure delay. However, predicting the performance of the future real time controllers (RTCs) that will be launched in space in the next decade is difficult. Their performance depend on numerous factor such as the electric consumption and supply, the size and efficiency of the processors, and the required RAM to make the calculations. If on-board WS\&C seems to be infeasible today with LUVOIR\cite{Pogorelyuk2021}, future space-based facilities might rely on current work carried in the context of the extremely large telescopes. Figure~\ref{fig:servo_lag} shows the impact of servo-lag on the obtained contrast level, with respect to the stellar magnitude in the case where sample C wavefront is corrected through predictive control, at both 100~Hz and 1000~Hz.

\begin{figure}[t]
    \centering
	\includegraphics[width=12cm]{Illustrations/Servo_lag-eps-converted-to.pdf}
	\caption{Obtained contrast level with respect to the stellar magnitude and the servo lag error for (top panel) 100~Hz and (bottom panel) 1000~Hz AO loop frequencies. The input aberrations are sample C.} % Figure caption
	\label{fig:servo_lag} % Label for referencing with \ref{bear}
\end{figure}
\end{comment}
        
\subsubsection{Random segment phasing error}
\begin{figure}[t]
    \centering
	\includegraphics[width=10cm]{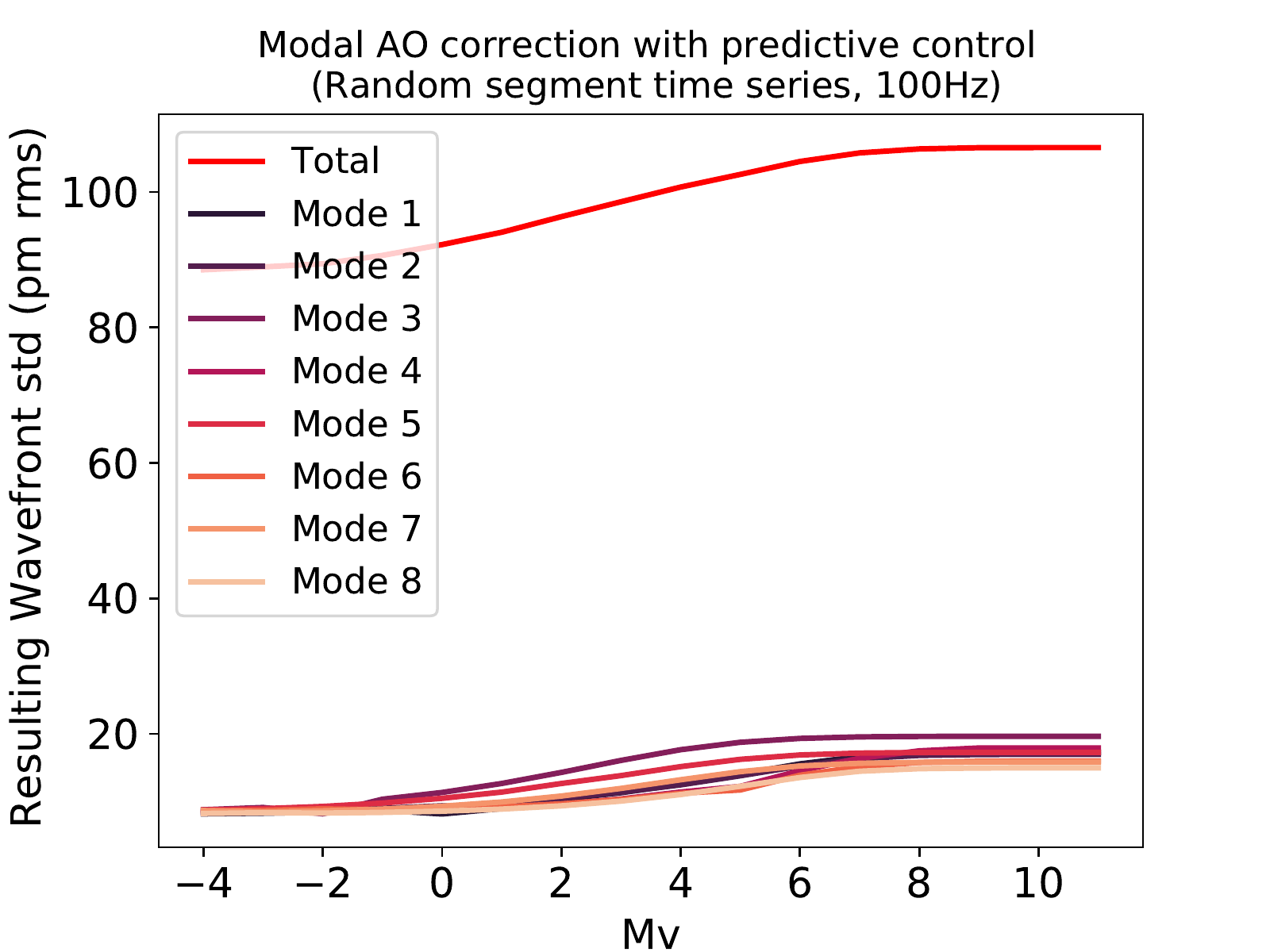}
	\caption{Correction of a random segment phasing error of 110pm rms over the pupil. The standard deviation of the residual aberrations is plotted with respect to the stellar magnitude. We assume a predictive control algorithm running at 100Hz and a WFS sampling of 64 pixels across the beam.} % Figure caption
	\label{fig:VariancevsMagnitude_RS} % Label for referencing with \ref{bear}
\end{figure}

The efficiency of the WS\&C algorithm described above depends strongly on the PCA decomposition offered by the structural time series. Indeed, most of the variance is contained in just a few modes, which assists the WFS measurement thanks to a higher signal-to-noise ratio for these few modes. We can however imagine a worst-case scenario where each segment is displaced randomly with respect to its neighbors. For the sake of illustration, we create a new time series with random pistons aberration for each of the 120 segments. The total standard deviation is 110~pm rms (close to sample C for comparison) and each segment carries 1/120th of the total variance. There is no particular vibrations in the segment basis and the temporal PSD is proportional to $f^{-1}$. We follow the recipe describe in this paper to correct for this time series. We first apply a PCA, but since each segment movement is random, the variance is distributed on many modes. The 10 first modes represent only 31.5\% of the total variance. In order to correct for a larger part of the signal in this simulation, we need to consider the 40 first modes that represent 70\% of the total. The 80 other modes left over carry individually less than 1\% of the total variance. In this particular time series, the signal to correct is lower than in the previous case. Also, the WFS noise is higher. Indeed, in contrast to the OPD simulation time series, there is no noticeable structures in the principle components. With that respect, it is harder for the WFS to distinguish between two separate modes. This induces stronger cross correlations between the modes and higher diagonal terms in the matrix $(GG^T)^{-1}$. This can be mitigated with a higher spatial resolution with the inconvenience to increase the photon noise per pixel. We plot in Fig.~\ref{fig:VariancevsMagnitude_RS} the standard deviation of the residuals, after the use of a predictive control algorithm running at 100~Hz, versus the stellar magnitude for such a time series. It shows a dramatic result where the wavefront is not sufficiently corrected, even with the use of a bright laser guide star. This result shows the structures that can be damped by optimizing the design of the telescope are also the easiest modes to correct with the on-board AO system. Therefore, these structures can be minimized on the sole condition that it does not introduce new modes spread randomly on more segments, even with if the level of optical aberration is globally decreased.

\section{Conclusion}
LUVOIR is a mission concept that aims to detect numerous exoEarths. However, its current mechanical structure and payload, including the large segmented primary mirror may introduce quickly varying wavefront aberrations in the coronagraphic instrument. These wavefront errors have tight requirements of $\sim$10~pm~RMS to achieve contrasts of $10^{-10}$ that will enable the detection and characterization of Earth-like planets. We considered a range of possibilities to minimize the aberrations upstream the coronagraph. First, the telescope dynamic aberrations may be optimized by damping some structural modes, or with the use of primary mirror and telescope control mechanisms, to reach a level of aberration at least equivalent to 10~pm rms (e.g. sample B above). In the future, an investigation about the feasibility of such a telescope stability will be performed by industrial partners. If 10~pm~RMS of aberration happens to be unrealizable, ECLIPS could be equipped with a fast high-order WS\&C system to filter part of this signal. Because of the lack of photons required to get a picometer-level sensitivity in a reasonable amount of time with a natural guide star, it might be required to associate the WS\&C architecture with a separated out-of-band laser guide star located in the line of sight or with a system of internal laser metrology. In any case, thanks to the steady conditions offered by the space environment, predictive control algorithms should be able to help correct OPD errors due to strong vibrations. The progress of the ground-based community, particularly regarding adaptive optics in the context of the future extremely large telescopes, could also fill some technological gaps.

\acknowledgments % equivalent to \section*{ACKNOWLEDGMENTS}       
The authors would like to thank Eric Gendron, Arielle Betrou-Cantou, and Jared Males for fruitful discussions.
The research was carried out at the Jet Propulsion Laboratory, California Institute of Technology, under a contract with the National Aeronautics and Space Administration (80NM0018D0004). Integrated Modeling of LUVOIR was performed by Lockheed Martin under a contract with the National Aeronautics and Space Administration (80MSFC20C0017).

% References
\bibliography{bib_GS}   %>>>> bibliography data in report.bib
\bibliographystyle{spiebib} % makes bibtex use spiebib.bst

\end{document}